\documentclass[%
 reprint,
 amsmath,
 amssymb,
 aps,
 pra
]{revtex4-2}

\usepackage{ragged2e}
\usepackage{caption}
\usepackage{babel}
\usepackage{subcaption}
\usepackage{graphicx}% Include figure files
\usepackage{dcolumn}% Align table columns on decimal point
\usepackage{bm}% bold math
\usepackage{inputenc}
\usepackage{lipsum}
\usepackage{xcolor}
\usepackage{tikz}
\usepackage{units}
\usepackage{diagbox}
\usepackage{multirow}

\usetikzlibrary{arrows.meta, calc, decorations.pathmorphing, shapes, positioning} 

\makeatletter
\pgfdeclaredecoration{single loop coil old}{final}{
  \state{final}[width=1pt]{
    \pgfpathmoveto{\pgfpointorigin}
    \pgfpathlineto{\pgfpoint{0.5*\pgfdecoratedpathlength-0.5*\pgfdecorationsegmentamplitude}{0pt}}
    \pgfpathcurveto
      {\pgfpoint{0.5*\pgfdecoratedpathlength + 0.25*\pgfdecorationsegmentamplitude}{0pt}}
      {\pgfpoint{0.5*\pgfdecoratedpathlength + 0.25*\pgfdecorationsegmentamplitude}{0.5*\pgfdecorationsegmentamplitude}}
      {\pgfpoint{0.5*\pgfdecoratedpathlength}{0.5*\pgfdecorationsegmentamplitude}}
    \pgfpathcurveto
      {\pgfpoint{0.5*\pgfdecoratedpathlength - 0.25*\pgfdecorationsegmentamplitude}{0.5*\pgfdecorationsegmentamplitude}}
      {\pgfpoint{0.5*\pgfdecoratedpathlength - 0.25*\pgfdecorationsegmentamplitude}{0pt}}
      {\pgfpoint{0.5*\pgfdecoratedpathlength + 0.25*\pgfdecorationsegmentamplitude}{0pt}}
    \pgfpathlineto{\pgfpoint{\pgfdecoratedpathlength}{0pt}}
  }
}

\pgfdeclaredecoration{single loop coil}{initial}{
  \state{initial}[width=0pt,next state=final]{
        \pgfpathmoveto{\pgfpointorigin}
        \pgfpathlineto{\pgfpoint{0.5*\pgfdecoratedpathlength-1.5*\pgfdecorationsegmentamplitude}{0pt}}
        \pgftransformxscale{2}   % scale x-direction
        \pgfpatharc{-90}{0}{\pgfdecorationsegmentamplitude}
        \pgftransformxscale{1}   % scale x-direction
  }
  \state{final}{
        \pgfpatharc{0}{180}{0.5*\pgfdecorationsegmentamplitude}
        \pgftransformxscale{2}   % scale x-direction
        \pgfpatharc{180}{180+90}{\pgfdecorationsegmentamplitude}
        \pgftransformxscale{1}   % scale x-direction
        \pgfpathlineto{\pgfpointdecoratedpathlast}
  }
}

\newcommand{\drawDetector}{
  \pgfpathellipse{\pgfpoint{0cm}{0cm}}{\pgfpoint{0.25 cm/4}{0}}{\pgfpoint{0}{0.25 cm/4*sin(15)}}
  \pgfusepath{stroke,fill}  

  \color{black}
  \pgfpathmoveto{\pgfpoint{0.25cm}{0cm}}
  \pgfpatharc{0}{180}{0.25cm}
  \pgfpatharc{180}{0}{0.25 cm and {0.25 cm*sin(15)}}
  \pgfusepath{stroke, fill}

  \pgfpathellipse{\pgfpoint{0cm}{0cm}}{\pgfpoint{0.25 cm}{0}}{\pgfpoint{0}{0.25 cm*sin(15)}}
  \pgfusepath{stroke}
}

\pgfdeclareshape{Detector_shape}{
  \anchor{center}{\pgfpoint{0cm}{0cm}}
  \anchor{north}{\pgfpoint{0cm}{0.25 cm}}
  \anchor{south}{\pgfpoint{0cm}{0cm}}
  \backgroundpath{
    \drawDetector    
  }
}

\pgfdeclareshape{BSM_shape}{
  \anchor{center}{\pgfpoint{0cm}{0cm}}
  \anchor{north}{\pgfpoint{0cm}{sin(45) * 0.5 cm}}
  \anchor{north west}{\pgfpoint{-cos(45) * 0.8 cm}{sin(45) * 0.75 cm + 0.25 cm}}
  \anchor{north east}{\pgfpoint{cos(45) * 0.8 cm}{sin(45) * 0.75 cm + 0.25 cm}}
  \anchor{south}{\pgfpoint{0cm}{-sin(45) * 0.5 cm}}
  \anchor{south west}{\pgfpoint{-cos(45) * 0.25 cm}{-sin(45) * 0.25 cm}}
  \anchor{south east}{\pgfpoint{cos(45) * 0.25 cm}{-sin(45) * 0.25 cm}}
  
  \backgroundpath{
    \pgfscope
      \pgftransformshift{\pgfpoint{-cos(45) * 0.75 cm}{sin(45) * 0.75 cm}}
      \pgftransformrotate{45}
      \drawDetector
    \endpgfscope
    \pgfscope
      \pgftransformshift{\pgfpoint{cos(45) * 0.75 cm}{sin(45) * 0.75 cm}}
      \pgftransformrotate{-45}
      \drawDetector
    \endpgfscope

    \pgfpathmoveto{\pgfpoint{-cos(45) * 0.75 cm}{sin(45) * 0.75 cm}}
    \pgflineto{\pgfpoint{0}{0}}
    \pgflineto{\pgfpoint{cos(45) * 0.75 cm}{sin(45) * 0.75 cm}}
    \pgfusepath{stroke}

    \pgfsetstrokecolor{black}
    \pgfsetfillcolor{gray!30}
    \pgfpathmoveto{\pgfpoint{0 cm}{sin(45)* 0.5 cm}}
    \pgflineto{\pgfpoint{cos(45)* 0.5 cm}{0 cm}}
    \pgflineto{\pgfpoint{0 cm}{-sin(45)* 0.5 cm}}
    \pgflineto{\pgfpoint{-cos(45)* 0.5 cm}{0 cm}}
    \pgfpathclose
    \pgflineto{\pgfpoint{0 cm}{-sin(45)* 0.5 cm}} 
    \pgfusepath{fill, stroke}
  }
}

\pgfdeclareshape{Qubit_shape}{
  \anchor{center}{\pgfpoint{0cm}{0cm}}
  \anchor{north}{\pgfpoint{0cm}{0.5 cm}}
  \anchor{south}{\pgfpoint{0cm}{-0.5 cm}}
  \anchor{east}{\pgfpoint{0.25 cm}{0 cm}}
  \anchor{west}{\pgfpoint{-0.25 cm}{0 cm}}
  \backgroundpath{
    
    \pgfscope
      \color{red!70!black}
      \pgftransformrotate{15}
      \pgfsetarrowsend{Latex[length=3.5pt]}
      \pgfpathmoveto{\pgfpoint{0 cm}{-0.25 cm}}
      \pgfpathlineto{\pgfpoint{0 cm}{0.3 cm}}
      \pgfusepath{stroke}
      \pgfpathcircle{\pgfpoint{0 cm}{0 cm}}{0.05 cm}
      \pgfusepath{fill}
    \endpgfscope
    \pgfsetstrokecolor{black}
    \pgfpathellipse{\pgfpoint{0 cm}{0 cm}}{\pgfpoint{0.25 cm}{0 cm}}{\pgfpoint{0 cm}{0.5 cm}}
    \pgfusepath{stroke}
  }
}
\makeatother

\tikzset{
    QM/.style = {rectangle, draw, inner sep=2pt},
    PPS/.style = {cylinder, draw, inner sep=2pt, shape aspect = 0.25},
    BSM/.style = {draw, shape = BSM_shape, scale = 0.6},
    PS/.style = {draw, circle, inner sep = 0.1 pt},
    Det/.style = {draw, shape = Detector_shape},
    entangled/.style = {
     decorate,
         decoration={single loop coil, amplitude= 0.3 cm}
        },
    signal/.style = {red!50!black},
    idler/.style = {blue!50!black},
    drive/.style = {green!50!black},
    pics/Pulse/.style={code={
        \filldraw[domain=-0.5:0.5,smooth, scale = 0.6, draw = #1, fill = #1, fill opacity = 0.5] plot  (\x,{exp(- \x*\x/0.05)});
    }},
    cw/.pic = {
        \filldraw[domain=-0.75:0.75, scale = 0.6, draw = red!50!black, fill = red!50!black, fill opacity = 0.5] plot  (\x,{exp(- abs(\x)/0.1)});
        \draw[scale = 0.6, opacity = 0.7] (0,-0.2) -- (0,1.1) node[at end, opacity = 1, above] {$t_c$};
        \draw[scale = 0.6, dashed, opacity = 0.7] (-0.4,1) -- (-0.4,-0.2);
        \draw[scale = 0.6, opacity = 0.7 ] (-0.4,-0.2) -- (0.4,-0.2)  node[midway, opacity = 1, below] {$T$} ;
        \draw[scale = 0.6, dashed, opacity = 0.7] (0.4,-0.2)  -- (0.4,1);
    },
    cwbf/.pic = {
        \fill[fill = blue!50!black, fill opacity = 0.5,scale = 0.6] (-0.5,0) rectangle (0.5,0.2);
        \draw[idler, scale = 0.6] (-0.5,0.2) -- (0.5,0.2);
    }
}
\usepackage{enumerate}
\newcommand{\hc}{\text{h.c.}}
\makeatletter
\def\l@en{\l@english}
\makeatother
\usepackage{physics}
\begin{document}
\preprint{APS/123-QED}
\title{The impact of multimode sources on DLCZ type quantum repeaters}

\author{Emil R. Hellebek}
% \author{Klaus Mølmer (?)}
\author{Anders S. Sørensen}
\affiliation{Center for Hybrid Quantum Networks (Hy-Q), Niels Bohr Institute, University of Copenhagen, Jagtvej 155A, Copenhagen DK-2200, Denmark}

\date{\today}

\begin{abstract}
    Long distance entanglement generation at a high rate is a major quantum technological goal yet to be fully realized, with the promise of many interesting applications, such as secure quantum computing on remote servers and quantum cryptography. One possible implementation is using a variant of the DLCZ-scheme by combining atomic-ensemble memories and linear optics with spontaneous parametric down conversion (SPDC) sources. As we edge closer to the realization of such a technology, the complete details of the underlying components become crucial. In this paper we consider the impact of the multimode emission from the SPDC source on quantum repeaters based on the DLCZ-scheme. We consider two cases, driving the SPDC using short Gaussian pulses and continuously. For pulsed driving, we find that the use of very narrow laser pulses to drive SPDC source is crucial to obtain high fidelity end-to-end entangled states but this puts demands on the peak intensity. By introducing a maximally allowed laser intensity, we find optimal pulse widths for each swap depth. For continuous driving, we find the temporal acceptance window of clicks relative to the heralding time to be a crucial parameter, and we can similarly optimize the acceptance window for each swap depth. For both cases, we thus identify optimal parameters given experimental limitations and aims. We have thus provided helpful knowledge towards the realization of long distance entanglement generation using the DLCZ-scheme.
\end{abstract}
\maketitle

\section{Introduction}
A long-distance quantum network, a quantum internet, is one of the central pillars of emerging quantum technologies \cite{kimbleQuantumInternet2008,wehnerQuantumInternetVision2018}. It would enable a host of exciting quantum applications such as remote blind quantum computing \cite{broadbentUniversalBlindQuantum2009,fitzsimonsPrivateQuantumComputation2017}, quantum key distribution \cite{abruzzoQuantumRepeatersQuantum2013,jingSimpleEfficientDecoders2021} and enhanced sensing \cite{wasilewskiQuantumNoiseLimited2010} on a larger scale. Long-distance quantum networks rely on the generation of long-distance entanglement through quantum repeaters \cite{briegelQuantumRepeatersRole1998,duanLongdistanceQuantumCommunication2001,sangouardQuantumRepeatersBased2011}. Here the total distance is partitioned into smaller segments, seperated by repeater stations containing one or more quantum memories (QMs). Entanglement between the neighboring stations is generated in parallel, and after a successful entanglement generation the entangled states are stored in the QMs. After neighboring pairs have generated entanglement, the entanglement is swapped to cover longer distances. A number of proposed candidates for quantum repeaters have been suggested, including using satellites \cite{liorniQuantumRepeatersSpace2021}, single trapped ions \cite{sangouardQuantumRepeatersBased2009}, color centers in diamond \cite{childressFaultTolerantQuantumCommunication2006} and quantum dots \cite{simonQuantumCommunicationQuantum2007}. 

One of the prime contenders is the DLCZ protocol proposed in Ref. \cite{duanLongdistanceQuantumCommunication2001}. The DLCZ protocol is based on the use of atomic ensemble quantum memories and linear optics to swap the entanglement for longer distances \cite{sangouardQuantumRepeatersBased2011}. Since the release of the DLCZ protocol a number of different improvements have been published including alternate detection setups for entanglement generation and swapping \cite{chenFaulttolerantQuantumRepeater2007,zhaoRobustCreationEntanglement2007,jiangFastRobustApproach2007}, the use of external sources with the incorporation of multiplexing \cite{sangouardLongdistanceEntanglementDistribution2007,simonQuantumRepeatersPhoton2007}, and swapping to other platforms allowing more advanced processing \cite{guHybridQuantumRepeaters2024,cussenotUnitingQuantumProcessing2025,tissotHybridSingleIonAtomicEnsemble2025}. In this work we focus on the scheme proposed by C. Simon \emph{et. al.} in Ref. \cite{simonQuantumRepeatersPhoton2007}, where the entanglement is generated by having two spontaneous parametric down conversion (SPDC) sources, with each source sending one photon into a QM and one photon towards a Bell-State measurement (BSM). A successful detection in the BSM creates entanglement between the photons stored in the two QMs. A key advantage of this protocol is that it is directly compatible with temporal multiplexing. By incorporating multimode memories, the scheme thus allows a significant boost to the quantum communication speed. 

To realize the full protocol in practice, it is essential to have a detailed understanding of the possible imperfections and operating condition of the protocol. Crucial work has been done for optimizing the QMs to allow for storage of multiple modes \cite{afzeliusMultimodeQuantumMemory2009}. Recent experimental work has also integrated the SPDC sources with QMs \cite{rielanderCavityEnhancedTelecom2016,lago-riveraTelecomheraldedEntanglementMultimode2021,rakonjacEntanglementTelecomPhoton2021,liuHeraldedEntanglementDistribution2021,busingerNonclassicalCorrelations12502022,hanniHeraldedEntanglementOndemand2025}. However, some issues with the SPDC sources are still outstanding as multiple pairs can be emitted and the multimode nature of the emitted state leads to infidelity \cite{braunsteinSqueezingIrreducibleResource2005,quesadaPhotonPairsNonlinear2022,hellebekCharacterizationMultimodeNature2024}. In a recent work, Ref. \cite{hellebekCharacterizationMultimodeNature2024}, we examined the multi-modality of the SPDC source and how this impacts the applicability as heralded single photon sources. In this paper, we extend this analysis and examine how the performance of a DLCZ-type quantum repeater is affected by the multi-modality of the sources, especially its impact on achievable rates and fidelities. We will furthermore find parameters that optimize the long distance entanglement generation rate for two different experimental settings: short pulse driving and (quasi) continuous driving. 

The paper is structured as follows: In Section \ref{sec:setup}, we describe the architecture of the protocol considered, and the operations for generating entanglement, swapping entanglement and post selection of the output.
In Section \ref{sec:pulse}, we investigate the case when driving the sources with short pulses.
In Section \ref{sec:cont}, we investigate the case when driving the sources continuously.
In Section \ref{sec:conc}, we conclude on the study, and discuss further possible investigations.

\begin{figure*}
    \centering
    \begin{tikzpicture}
        \node at (-4,0.45) {\textbf{a)}};
        \node[BSM, color = blue!50!black] (BSM) at (0,0) {};
        \foreach \i/\j/\k/\l in {-1/south west/east/$A$,1/south east/west/$B$}{
            \node[PPS, rotate = 90] (PPS\i) at (\i*1.5,0) {SPDC};
            %\draw[drive, rotate = 90] (PPS\i) --++ (0.75,0);
            \node[QM] (QM\i) at (\i*3.75,-0.75) {QM\l};
            \node[QM] (QM2\i) at (\i*3.75,-1.75) {QM\l};
            \draw[signal] (QM\i.\k) --++(-\i*1.5,0) to [out = 90+\i*90, in = 270](PPS\i.west);
            \draw[idler, out = 270] (PPS\i.west)  to [in = 90-\i*90] ++ (-\i*0.5, -0.5) -- ++(-\i*0.5,0) to [out = 90+\i*90, in = 270+\i*45]  (BSM.\j);
            \draw pic at (\i*2.75,-0.75){Pulse={red!50!black}};
            \draw pic at ($(PPS\i.west)+(-\i*0.75,-0.5)$) {Pulse={blue!50!black}};
        }
        \node at (0,-1) {$\Downarrow$};
        \draw[entangled] (QM2-1) -- (QM21);
    \end{tikzpicture}
    \hfill
    \begin{tikzpicture}
        \node at (-4,0.45) {\textbf{b)}};
        \node[BSM, color=blue!50!black] (BSM) at (0,0) {};
        \foreach \i/\j/\k/\l in {-1/south west/east/$A$,1/south east/west/$B$}{
            \node[PPS, rotate = 90] (PPS\i) at (\i*1.5,0) {SPDC};
            %\draw[drive, rotate = 90] (PPS\i) --++ (0.75,0);
            \node[QM] (QM\i) at (\i*3.75,-0.75) {QM\l};
            \node[QM] (QM2\i) at (\i*3.75,-1.75) {QM\l};
            \draw[signal] (QM\i.\k) --++(-\i*1.5,0) to [out = 90+\i*90, in = 270](PPS\i.west);
            \draw[idler, out = 270] (PPS\i.west)  to [in = 90-\i*90] ++ (-\i*0.5, -0.5) --++(-\i*0.5,0) to [out =90+\i*90, in = 270+\i*45]  (BSM.\j);
            \draw pic at (\i*2.75,-0.75){cw};
            \draw pic at ($(PPS\i.west)+(-\i*0.75,-0.5)$) {cwbf};
        }
        \draw[thick, dashed, ->] (BSM.north west) --++ (0,0.6) -| (-2.75, 0.4);
        \node at (0,-1) {$\Downarrow$};
        \draw[entangled] (QM2-1) -- (QM21);
    \end{tikzpicture}

    \vspace{0.5 cm}

    \begin{tikzpicture}
        \node at (-3.25,0.25) {\textbf{c)}};
        \node[BSM, color = red!50!black] (BSM) at (0,0) {};
        % \draw[scale = 0.6, transform shape] pic at (BSM) {BSM={red!50!black}};
        \foreach \i/\j/\k/\l in {-1/south west/$B$/$A$, 1/south east/$C$/$D$}{
            \node[QM] (QMC\i) at (\i,-0.5) {QM\k};
            \node[QM] (QME\i) at (3*\i,-0.5) {QM\l};
            \node[QM] (QME2\i) at (3*\i,-1.5) {QM\l};
            \draw[signal, out = 90+90*\i, in = 270+\i*45 ] (QMC\i)  to (BSM.\j);
        }
        \draw[entangled] (QME-1) -- (QMC-1);
        \draw[entangled] (QMC1) -- (QME1);
        \node at (0,-0.8) {$\Downarrow$};
        \draw[entangled] (QME2-1) -- (QME21);
        
    \end{tikzpicture}
    \hfill
    \begin{tikzpicture}
        \node at (-3.75,1) {\textbf{d)}};
        \foreach \i/\k in {-1/A, 1/B}{
            \node[PS] (PS\i) at (\i*3,0) {$\sigma_i^\k$};
            
            \foreach \j/\l in {-1/L, 1/U}{
                \node[QM] (QM\i\j) at (\i*2,\j*0.3) {QM$\k_\l$};
                \node[Det, rotate = (-\i)*(90-\j*22.5), signal, scale = 0.75] (Det\i\j) at (\i*3.5,\j*0.3) {};
            }
        }
        \foreach \j/\k/\h in {-1/south east/south west/,1/north east/north west}{
            \draw[entangled] (QM-1\j) -- (QM1\j);
            \draw[signal, out = 180, in = \j*45 ] (QM-1\j)  to (PS-1.\k);
            \draw[signal, out = 0, in = 180-\j*22.5] (QM1\j)  to (PS1.\h);
            \draw[signal, out = 180-\j*45, in = -\j*22.5] (PS-1.\h) to (Det-1\j.center);
            \draw[signal, out = \j*45 , in = 180+\j*22.5] (PS1.\k) to (Det1\j.center);
        }
    \end{tikzpicture}
    \caption{
    \textbf{a)} and \textbf{b)} Schematics of the entanglement generation for pulsed driving \textbf{a)} and continuous driving \textbf{b)}. The idler photons (blue lines) from both SPDC sources are sent to a central BSM station, whereas the signal photons (red lines) are sent to the respective quantum memory. A detection will herald the entanglement between the light stored in the two QMs. In \textbf{b)} we indicate the post-processing, where a click have been registered at $t_c$, and the acceptance window $T$, indicates, how much of the light associated with that detection will be included later. \textbf{c)} Schematic of an entanglement swap. Two neighboring pairs of entangled QMs have been generated. The photons are read out from the central QMs and sent to a BSM based on photon number resolving (PNR) detectors. The detection of a single photon swaps the entanglement to the outer QMs. \textbf{d)} Schematic of the final readout and post-selection. Light from the four remaining QMs are read out. At both ends, a Pauli operator acts on the output light before being read by PNR detector. 
    }
    \label{fig:arch}
\end{figure*}
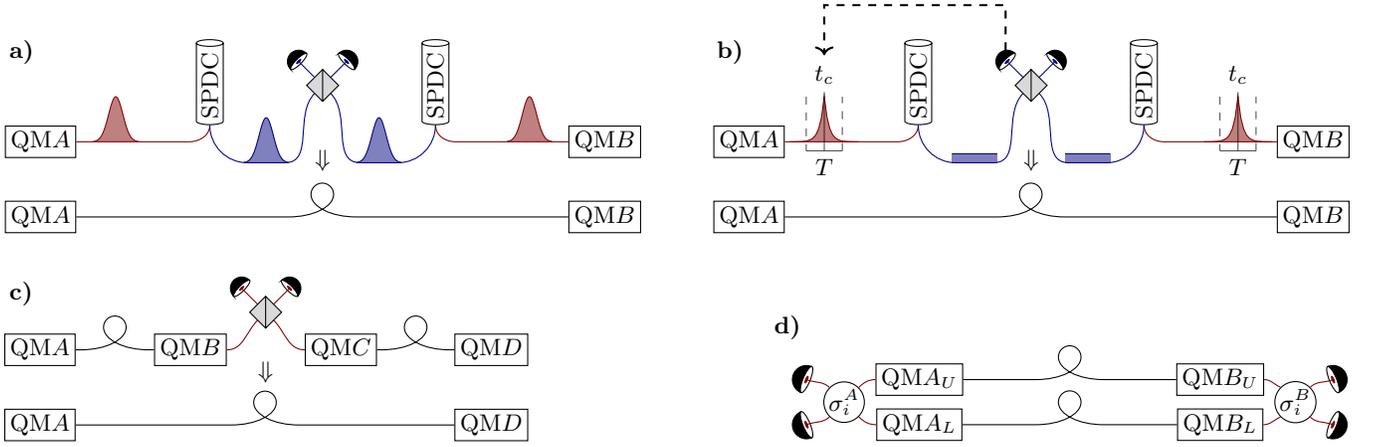
\section{Setup and protocols}\label{sec:setup}

In this section we present a short summary of the protocol: We will establish two parallel chains each containing $2^n$ elementary links of length $L_0=L/2^n$, where $n$ is the swap depth and $L$ is the total distance. Each repeater node consists of two SPDC sources, two QMs and a BSM-station used for entanglement swapping. Furthermore, we place a BSM-station halfway between repeater nodes used for entanglement generation. After the long-distance entanglement have been established in the two chains, we post-select to get one purified entangled pair. We will detail the exact operations for entanglement generation, entanglement swapping and readout below.

We will in this study examine two main driving fields for operating the sources: Driving the SPDC sources using short Gaussian pulses and driving the SPDC sources continuously. The driving fields will give rise to an interaction strength $\chi$ with the time dependence of the driving field. In our previous work (Ref. \cite{hellebekCharacterizationMultimodeNature2024}) we studied these two methods, and examined the multi-modality of the heralded state. In Appendix \ref{app:source}, we summarize how the source is modelled in this study.

When employing Gaussian pulses, which are narrow compared to the cavity decay time $1/\kappa$ to drive the SPDC, the state in the signal beam is heralded upon detection in the idler beam. In this case the main parameter controlling the multimode nature of the source is the width of the Gaussian pulse $\sigma$, and we found that both the single-photon purity and the multi-photon emission probability decrease when increasing $\sigma$.
On the other hand, when the source is driven continuously, a detection in the idler beam at $t_c$, heralds a photon in the signal beam around this time. We, thus, set an acceptance window $T$ symmetrically around $t_c$ to capture the signal photon. In this case increasing $T$ increases the probability to have any number of photons, i.e. both the efficiency of extracting single photons and the erroneous probability to collect multiple photons. Finally, if we have captured multiple photons, the probability that the photons occupy orthogonal modes increase as $T$ increases. 

\subsection{Entanglement generation}
The entanglement generation process is sketched in fig. \ref{fig:arch}\textbf{a)} and \textbf{b)} for pulsed and continuous driving, respectively. To generate entanglement two SPDC sources in neighboring nodes are driven. We assume the source is driven sufficiently weak to only emit up to two photon pairs. The full multimode state is thus given by \cite{hellebekCharacterizationMultimodeNature2024}

\begin{widetext}
\begin{align}
    \ket{\Psi_{A}} = \pqty{\sqrt{P_0} + \sqrt{P_1}\sum_\ell\sqrt{w_\ell}a_{\ell,s}^\dagger a_{\ell,i}^\dagger + P_1 \sum_{k\ell} \frac{\sqrt{w_iw_j}}{2}  a_{\ell,s}^\dagger a_{k,s}^\dagger a_{\ell,i}^\dagger a_{k,i}^\dagger}\ket{\emptyset},
\end{align}    
\end{widetext}
where $A$ denotes the SPDC source, $\ell$ denotes a specific mode of the eigen-decomposition with weight $w_\ell$, which can be calculated from the drive $\chi(t)$. $a_{\ell,s}$ ($a_{\ell,i}$) is the annihilation operator of the signal (idler) photon of mode $\ell$, and $P_1$ is the probability of the source emitting one photon pair.

The output from each source is split using a dichroic mirror, such that the light in the signal beam is stored in the QM, while the light in the idler beam is sent over a long distance to the central BSM station. Here, the beams from each node are mixed on a beamsplitter, before the beams impinge on single-photon detectors. A detection in any of the two detectors heralds the entanglement between the light in the QM in each node 
% COMMENT ON MEMORIES BEING ENTANGLED
\begin{align}\label{eq:egen}
    \rho_{0,AB|t_c} = \frac{\Tr_{i}\bqty{b_{+i}(t_c)\ketbra{\Psi_{A}}\ketbra{\Psi_{B}}b_{+i}^\dagger(t_c)}}{\ev{b_{+i}^\dagger(t_c) b_{+i}(t_c)}},
\end{align}
where $\Tr_{i}[\cdot]$ is the trace over idler photons from both memories, $b_{\pm i}(t_c) = (a_i(t_c) \pm b_i(t_c))/\sqrt{2}$ are the annihilation operators associated with each detector, and we have assumed the detection to be in the "+"-mode. The detection is modeled using annihilation operators. While this treatment is strictly speaking incompatible with measurement theory (see eg. Ref. \cite{bergouQuantumStateDiscrimination2007}), it is a good assumption when the state suffers from high losses, which is the case for the long transmission between repeater node and BSM-station. For the pulsed case, we will average over all detection times 
\begin{align}\label{eq:pulse_op}
    \rho_{0,AB}^\text{pulsed} = \frac{\int_\mathbb{R} dt_c \ev{b_{+i}^\dagger(t_c) b_{+i}(t_c)}\rho_{0,AB|t_c}}{\int_\mathbb{R} dt \ev{b_{+i}^\dagger(t) b_{+i}(t)}}.
\end{align}

When employing continuous driving, we will introduce an acceptance window chosen symmetrically around $t_c$, and disregard all photons outside the acceptance window, as illustrated in Fig. \ref{fig:arch}\textbf{b}). We thus have the heralded state
\begin{align}\label{eq:cw_op}
    \rho_{0,AB|t_c,T}^\text{cont} = \Tr_{\neg T}\bqty{\rho_{0,AB|t_c}},
\end{align}
where $\Tr_{\neg T}[\cdot]$ denotes the trace over times outside the interval $[t_c-T/2,t_c+T/2]$.

The probability to successfully generate the elementary link in one attempt, $\mathbf{P}_0$, is given by
\begin{align}
    \mathbf{P}_0=2\eta_{LD}\int_\mathcal{T} dt_c\,\ev{b_+^\dagger(t_c)b_+(t_c)},
\end{align}
where $\mathcal{T}$ is the total time looking for photons and $\eta_\text{LD} = \eta_d \exp(-L_0/2L_\text{att})$, is the efficiency from the node to the heralding station, with $\eta_d$ the detection efficiency and $L_\text{att}$ the attenuation length, set to $\unit[22]{km}$. For the pulsed driving, we will let $\mathcal{T}$ cover the whole time range, hence the integral will become one, and the success probability reduces to
\begin{align}
    \mathbf{P}_0^\text{pulsed}=2\eta_{LD}P_1\bqty{1+P_1\pqty{1+\varpi_1}},
\end{align} 
where $\varpi_1$ is the single photon purity, i.e. $\varpi_1=\sum_i w_i^2$.

For continuous driving, we require that the acceptance window is much shorter than the time spent looking for photons, i.e. $T\ll\mathcal{T}$. Here the success probability can be calculated directly from the Bogoluibov transformation \cite{hellebekCharacterizationMultimodeNature2024}
\begin{align}
    \mathbf{P}_0^\text{cont}=4\eta_{LD}\chi^2\pqty{1+4\frac{\chi^2}{\kappa^2}} \mathcal{T} ,
\end{align}
where $\kappa$ is the FWHM linewidth of the cavity around the SPDC source.

Once we obtain a successful heralding detection the stored photon is entangled between the two memories. We will assume that the source is driven sufficiently weak to only include terms in the density matrix up to linear order in the intensity of the pulse. This allows us to write the density matrix for any swap depth $n$ using a closed set of coefficients
\begin{widetext}
\begin{align}\label{eq:state}
    \rho_{n,AB} = c_{00,n} \rho_{00}
    + c_{10,n}\rho_{10}
    + c_{11,n} \rho_{11}
     + c_{20,n}\rho_{20}
    + \rho_{1001,n}
    + \rho_{2011,n},
\end{align}

where $\rho_{kl}$ is the normalized diagonal density matrix element with $k$ photons in memory $A$ and $l$ photons in memory $B$, $c_{kl,n}$ is the coefficient associated with the diagonal density matrix element $\rho_{kl}$ at swap depth $n$ and $\rho_{ijkl,n}$, is the off-diagonal element in matrix between $i$ photons in memory $A$ and $j$ photons in memory $B$ and $k$ photons in the memory $A$ and $l$ photons in memory $B$ at swap depth $n$. 

\subsection{Entanglement swap}
When two neighboring entanglements of the same swap depth, $n$, have successfully been established, we attempt entanglement swapping to reach swap depth $n+1$. The entanglement swapping is sketched in Fig. \ref{fig:arch}\textbf{c)}. The light stored in QM$B$ and QM$C$ is read out with a combined memory efficiency $\eta_m$. 
This light is then sent to a BSM, with number resolving single-photon detectors with efficiency $\eta_d$. We can thus introduce a single loss channel, such that the annihilation operators of the stored light undergo the transformation $b_s(t)\to \sqrt{\eta_m\eta_d} b_d(t) + \sqrt{1-\eta_m\eta_d}b_\perp(t)$, where $b_d$ is the annihilation operator associated with the detector channel from QM$B$ and $b_\perp$ is the annihilation operator of the loss channel. The light from the quantum memories $B$ and $C$ are mixed evenly on a beamsplitter before the detectors, such that $b_{\pm d}(t) = (b_d(t) \pm c_d(t))/\sqrt{2}$. A successful heralding occurs when one of the two detectors find exactly one photon, yielding the transformations 
\begin{align}\label{eq:eswap}
    \rho_{n+1,AD} = \frac{\int dt_s \ev{b_+(t_s)\Tr_\perp[\rho_{n,AB}\otimes\rho_{n,CD}]b^\dagger_+(t_s)}{\emptyset_{b_{\pm d}}}}{\int dt'_s \Tr[\ev{b_+(t'_s)\rho_{n,AB}\otimes\rho_{n,CD}b^\dagger_+(t'_s)}{\emptyset_{b_{\pm d}}}]},
\end{align}
where $\ket{\emptyset_{b_{\pm d}}}$ is the state with no photons in the detector channels. We have chosen the single-click entanglement swapping protocol of the DLCZ-scheme, as this is easier to implement and yields the highest rate at low swap depths. Furthermore, this assumption allows us to investigate the influence of the imperfections of the source for the simplest possible scenario.

The transformation of $\rho_n$ to $\rho_{n+1}$ yields a set of recursive equations for the coefficients in Eq. \eqref{eq:state}. The initial conditions are obtained by combining Eq. \eqref{eq:egen} with Eq. \eqref{eq:pulse_op} when using short pulses and Eq. \eqref{eq:cw_op} when using continuous driving. In Appendix \ref{app:pulse} and \ref{app:CW} we give the full description of the density matrix for pulsed and continuous driving and express the recursive equations directly. 

The probability of a successful entanglement swap is
\begin{align}
    \mathbf{P}_{n+1} = 2\eta_m\eta_d\bqty{1-\eta_m\eta_d\pqty{\frac{c_{10,n}}{2}+c_{11,n}+c_{20,n}}+\eta_m^2\eta_d^2\frac{c_{20,n}}{2}}\bqty{\frac{c_{10,n}}{2}+c_{11,n}+\pqty{1-\eta_m\eta_d}c_{20,n}}.
\end{align}    

\subsection{Final readout and post-selection}
We sketch the final readout in Fig. \ref{fig:arch}\textbf{d)}. After creating two entangled pairs between the end nodes, the light from all four remaining QMs is retrieved. We then measure the light at each end in one of the different Pauli bases and post-select on a photon being measured at each end. This post-selection suppresses the unwanted vacuum component ($c_{00,n}$ in Eq. \eqref{eq:state}) unless mixed with the unlikely two-photon component ($c_{11,n}$ in Eq. \eqref{eq:state}). By setting a sufficiently low driving strength in the initial entanglement generation, we can thus obtain a high-fidelity entangled state. 

The probability of a successful post-selection can be found from the traces of the density matrix
\begin{align}
    \mathbf{P}_{\text{PS}|n} = 2\eta_m^2\eta_d^2\qty{c_{11,n}\bqty{1-\eta_m\eta_d\pqty{c_{10,n}+\pqty{2-\eta_m\eta_d}\pqty{c_{11,n}+c_{20,n}}}}+\bqty{\frac{c_{10,n}+2\pqty{1-\eta_m\eta_d}\pqty{c_{11,n}+c_{20,n}}}{2}}^2}.
\end{align}

\subsection{Fidelity and rate}
The desired state of the read-out photons after post-selection is $\ket{\Psi_{AB}}=\frac{1}{\sqrt{2}}\pqty{\ket{01}\pm\ket{10}}$, in the basis of left side, right side using the notation from above and where $\ket{0}$ ($\ket{1}$) represents a photon from the upper (lower) chain. 
Assuming, we get a detection pattern yielding the plus sign in the entanglement generation, we obtain the fidelity of final state
\begin{align}
    F &= \frac{(\bra{01}+\bra{10})\tilde{\rho}_n\otimes \tilde{\rho}_n(\ket{01}+\ket{10})}{2\mathbf{P}_{\text{PS}|n}}=\frac{1-\ev{\sigma_z^A\sigma_z^B}+\ev{\sigma_x^A\sigma_x^B}+\ev{\sigma_y^A\sigma_y^B}}{4\mathbf{P}_{\text{PS}|n}},
\end{align}
where the second equality shows how to find the fidelity from measurements of the Pauli-operators, $\tilde{\rho}_n$ is $\rho_n$ after readout of all stored photons and $\sigma_i^A$ ($\sigma_i^B$) is the Pauli-$i$ operator on the left (right) side, defined as 
\begin{align}
    \sigma_x = \sum_\ell (\ketbra{1_\ell \emptyset}{\emptyset 1_\ell}+\ketbra{\emptyset 1_\ell }{1_\ell \emptyset}) && 
    \sigma_y = -i \sum_\ell (\ketbra{1_\ell \emptyset}{\emptyset 1_\ell}-\ketbra{\emptyset 1_\ell }{1_\ell \emptyset}) &&
    \sigma_z = \sum_\ell (\ketbra{1_\ell \emptyset}{1_\ell \emptyset}-\ketbra{\emptyset 1_\ell }{\emptyset 1_\ell }),
\end{align}
in the basis where the first letter describe photons from the upper memory and the second photons from the lower memory. 
\newpage

\end{widetext}
The success probabilities for the entanglement generation, entanglement swap and post-selection can be 
combined into a rate of end-to-end entanglement generation $\tilde{R}$ pr. memory without multiplexing \cite{sangouardQuantumRepeatersBased2011}
\begin{align}
    \tilde{R}_n = \frac{c}{L_0}\frac{1}{2^{n+2}}\frac{\mathbf{P}_0\mathbf{P}_1\cdots \mathbf{P}_n \mathbf{P}_\text{PS}}{\pqty{3/2}^{n+1}},
\end{align}
where $c$ the speed of light in the fiber and the $2^{n+2}$ are the number of memories and the $(3/2)^{n+1}$ factor accounts for the establishment of pairs in parallel of entangled memories at all swap depths. 

In the limit of vanishing driving power and e.g. infinitely short pulse for the pulses, the desired state will be obtained with unit fidelity at zero rate. Higher rates decrease the fidelity but increase the rate. We can thus set a target fidelity, which will yield the driving power and the total end-to-end entanglement generation time. We will now do this for the two cases separately.

\section{Pulsed driving}\label{sec:pulse}

\begin{figure}
    \includegraphics[width=\columnwidth]{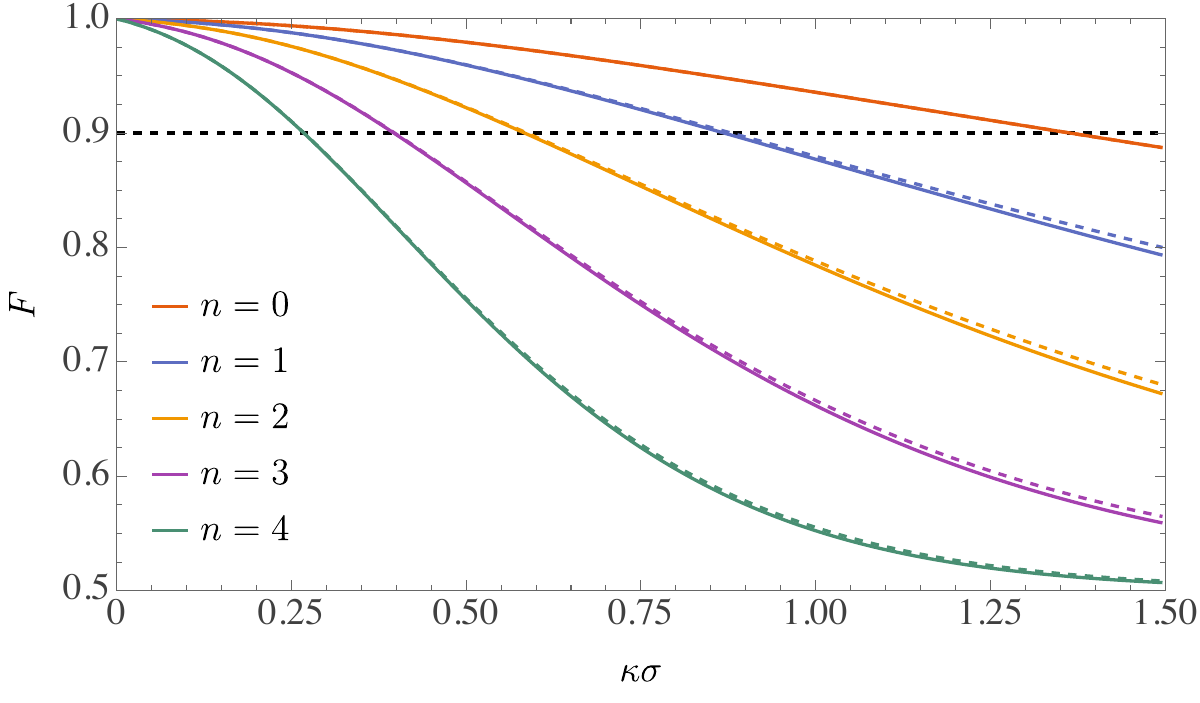}
    \caption{Fidelity due to the multimode nature. Full lines show the complete model, and dashed lines show the approximated version, from Eq. \eqref{eq:baseFid}. The horizontal dashed line shows 90\% fidelity, which is the target set for later. The lines go in descending order in $n$ when going up the $F$ axis.}
    \label{fig:Fidelity}
\end{figure}

In this section we investigate the results when driving the SPDC source with a narrow pulse. The SPDC source is embedded in a cavity, and we assume that the pulse will make a single pass through the SPDC. For simplicity, we will assume a Gaussian drive, i.e. the interaction between the driving pulse and the non-linear medium $\chi$ has the time-dependence  $\chi(t) = x/(\sqrt{2\pi}\sigma)\exp(-t^2/2\sigma^2)$, where $x$ is a dimensionless characterization of the driving strength. In Appendix \ref{app:pulse} we give a detailed description of the density matrix in this scenario.

As described above, the multimodality of output state from the SPDC source depends on the width of the Gaussian pulse $\sigma$ employed. An impure state initial state will lead to decreasing coherence after each swap. Conclusively, some infidelity is introduced even to zeroth order in the driving power 
\begin{align}\label{eq:baseFid}
    F = \frac{1+\sum_\ell w_\ell^{2^{n+1}}}{2}\approx \frac{1+\varpi_1^{2^{n}}}{2},
\end{align}
where $w_\ell$ are the weights of the single-photon mode decomposition and $\varpi_1$ is the single-photon purity defined as $\sum_\ell w_\ell^2$. This approximation is equivalent to assuming that the coherence terms is multiplied by a factor of $\varpi_1$ for each entanglement swap and in the post-selection. Using the results from Ref. \cite{hellebekCharacterizationMultimodeNature2024}, we can find the eigenmode decomposition of the emitted light using numerical methods. 

In Fig. \ref{fig:Fidelity} we plot the baseline fidelity, defined in the limit as $P_1\to 0$, against different pulse widths $\sigma$. In the small $\sigma$ limit, where the emitted state is pure the impact to fidelity is negligible. However, as $\sigma$ increase the underlying state becomes more impure and thus affects the fidelity of the final state. This is especially damaging for higher swap depth, as more swaps have to be performed. Furthermore, we note that, the approximation of Eq. \eqref{eq:baseFid} corresponds very closely to the actual values.

\begin{figure}
    \includegraphics[width=\columnwidth]{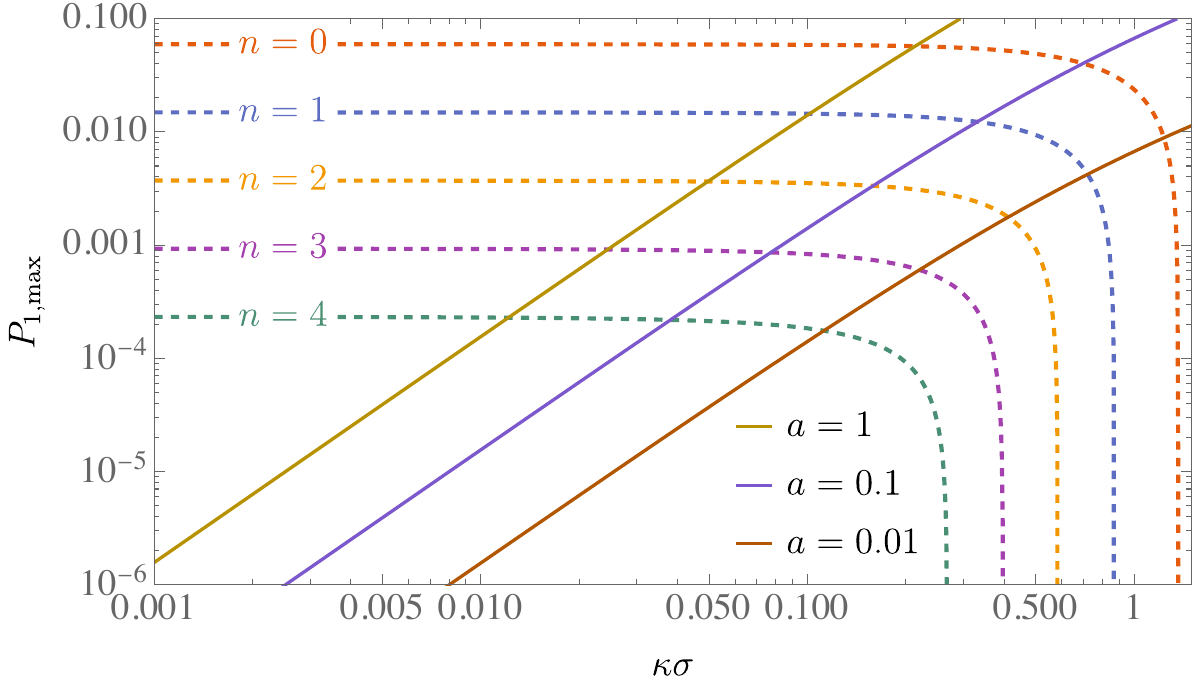}
    \caption{The full lines show the maximal value $P_1$ can attain, if the maximal intensity of the laser is $I_\text{max}=a I_\text{thres}$. The dashed lines show the target $P_1$ for different swap depths using a target fidelity of 90\%. With a source limited by a certain value of $a$, the ideal pulse width is found by the crossing of the dashed and full lines. We have used $\eta_d=0.9$ and $\eta_m=0.8$. The full lines go in ascending order in $a$ when going up the $P_{1,\text{max}}$ axis.}
    \label{fig:Threshold}
\end{figure}
\begin{table}
    \begin{tabular}{|c |c c | c c |}
        \hline $a$ & \multicolumn{2}{c|}{0.01} & \multicolumn{2}{c|}{0.1}  \\\hline
        $n$ & $\kappa\sigma^*$ &  $L\, [\unit{km}]$ & $\kappa\sigma^*$ &  $L\, [\unit{km}]$ \\\hline
        0 & 1.24 & 100-105 & 0.70 & 100-124 \\
        1 & 0.72 & 105 - 252 & 0.33 & 124 - 291 \\
        2 & 0.41 & 252 - 627 & 0.16 & 291 - 678 \\
        3 & 0.22 & 627 - 1512 & 0.08 & 678 - 1563 \\
        4 & 0.11 & 1512- & 0.04  & 1563- \\\hline\hline
        $a$ & \multicolumn{2}{c|}{1} & \multicolumn{2}{c|}{$\infty$} \\\hline
        $n$ & $\kappa\sigma^*$ &  $L\, [\unit{km}]$ & \multicolumn{2}{c|}{$L\, [\unit{km}]$} \\\hline
        0 & 0.21 & 100 - 133 & \multicolumn{2}{c|}{100 - 133}\\
        1 & 0.10 & 133 - 298 & \multicolumn{2}{c|}{133 - 299} \\
        2 & 0.05 & 298 - 683 & \multicolumn{2}{c|}{299 - 684} \\
        3 & 0.02 & 683 - 1568 & \multicolumn{2}{c|}{684 - 1568} \\
        4 & 0.01 & 1568- & \multicolumn{2}{c|}{1568-} \\\hline
    \end{tabular}
    \caption{For each combination of swap depth and maximally allowed intensities, using $I_\text{max}= a I_\text{thres}$, we show the optimal value for the pulse width $\kappa \sigma^*$ and the range of distances above $\unit[100]{km}$ and up to $n=4$ where a particular swap depth yields the highest rate. For $a\to \infty$ the ideal pulse width is $\sigma\to 0$. We have used a target fidelity of $90\%$, $\eta_d=0.9$, $\eta_m=0.8$ and $L_\text{att}=\unit[22]{km}$.}
    \label{tab:sigma*}
\end{table}
\begin{figure*}
    \includegraphics[width=\linewidth]{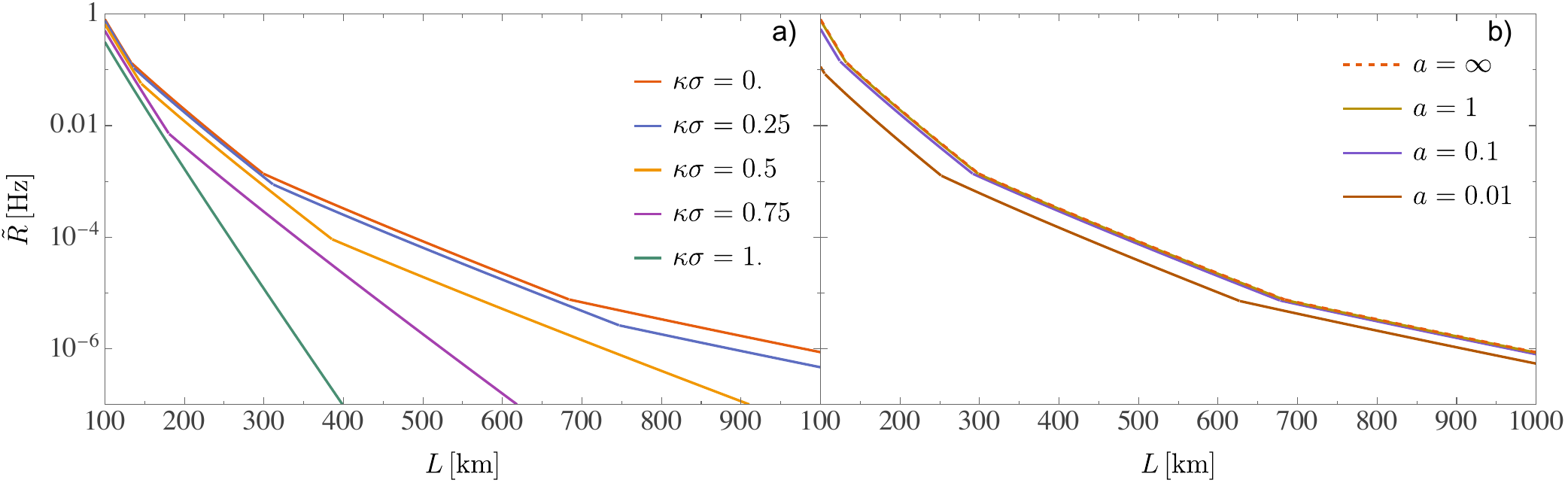}

    \caption{Un-multiplexed rates pr. quantum memory for a) fixed pulse widths $\sigma$ and b) fixed maximally allowed laser intensity, $a = I_\text{max}/I_\text{thres}$. Note that $a=\infty$ in b) corresponds to the $\kappa\sigma=0$ in a).  We have used a target fidelity of $90\%$, $\eta_d=0.9$, $\eta_m=0.8$, $c=\unitfrac[2\cdot10^5]{km}{s}$ and $L_\text{att}=\unit[22]{km}$. For a) (b)) the lines go in descending (ascending) order in $\kappa \sigma$ ($a$) when going up the $\tilde{R}$ axis.}
    \label{fig:rates_pulsed}
\end{figure*}
A target fidelity will thus effectively set an upper limit on the possible swap depth for a given pulse width, as the fidelity will only further decrease for a non-vanishing $P_1$.
We can translate the target fidelity into a target value of $P_1$ given $n$ and $\sigma$. In Fig. \ref{fig:Threshold} we plot the obtained value of $P_1$ using a target fidelity of 90\% for the different swap depths and pulse widths. However, a narrow pulse with a large value of $P_1$ will result in high peak driving intensities, which may represent a challenge for experiments. To investigate this, we set a maximal power for the experiments $I_\text{max} = a I_\text{thres}$, where $I_\text{thres}$ is the threshold power at which point the output of the SPDC source becomes exponentially enhanced (which corresponds to $\chi = \kappa/2$, see Appendix \ref{app:source} for details). We translate the maximal driving intensity into a value of $P_1$ (to the lowest order in $\chi(t)$) obtaining
\begin{align} \label{eq:P1max}
    P_{1,\text{max}} = a\frac{\pi \kappa^2\sigma^2}{2} e^{\kappa^2\sigma^2}\bqty{1-\erf\pqty{\kappa\sigma}},
\end{align}
where the detailed calculation is left for Appendix \ref{app:P1}.
Higher values of $a$ allows for more narrow pulses, and thus a higher possible value of $P_1$. We plot $P_{1,\text{max}}$ for three different values of $a$. Since, the rate increase with $P_1$, we choose the crossing of the lines set by  the target fidelity and Eq. \eqref{eq:P1max}. We report these pulse widths in Tab. \ref{tab:sigma*} for each value of $a$.

We should note that for very short drive pulses the output single photon mode is a decaying exponential in time \cite{hellebekCharacterizationMultimodeNature2024}. When employing cavity assisted quantum memories a photon with this mode shape has been shown to be particularly difficult to store \cite{kollath-bonigFastStoragePhotons2024}. The influence of the mode shape on the memory efficiency has not been included in the present study, as this effect will depend on the exact experimental details.

We can now calculate the rates for different transmission lengths and different swap depths. The rate is first maximized over the different swap depths after fixing the pulse width. The result is plotted in Fig. \ref{fig:rates_pulsed}a) for a few values of $\sigma$. We see that slower rates are obtained when increasing the pulse width, which is ultimately caused by the increased impurity of the emitted state. Furthermore, we note that for some pulse widths only the lowest swap depths are allowed. In Fig. \ref{fig:rates_pulsed}b) we fix the maximally allowed intensity and optimize the rate for a given length. The optimal pulse width $\sigma^*$, i.e. the lowest allowed pulse width for a given swap depth, is chosen according to Fig. \ref{fig:Threshold}. We see, unsurprisingly, that a lower allowed intensity yields a lower rate. However, the negative impact of increasing the swap depth is now smaller, as the pulse width can be decreased yielding a higher purity initial state, which decreases the impact on $P_1$. Furthermore, we observe that even a modest value of $a=1$ has essentially no impact on the obtainable rate. The influence of limited driving power, is particularly small at long distances since the DLCZ-protocol in itself require very weak driving for long distances. We report in Tab. \ref{tab:sigma*} the range of distances where each swap depth yields the highest rate given $a$. 

\section{Continuous driving} \label{sec:cont}
In this section, the impacts of driving the SPDC source with a continuous drive are investigated. For the entanglement generation, we will detect a heralding click at time $t_c$. We will only store photons inside the memory, if they are temporally close to the heralding click $t_c$. This is set by the length of the acceptance window $T$. $T$ needs to be chosen sufficiently long to allow the entire pulse to be extracted, but at the same time sufficiently short to avoid additional clicks from pairs generated at a different time. 

We will write the driving strength as $\chi(t) = x\kappa/2$, where $x$ is dimensionless. The heralded output state is pure to the lowest order in the driving intensity, but becomes multimode to first order. The detailed calculations describing the state at different swap depths are left for appendix \ref{app:CW}.

\begin{figure}
    \includegraphics[width=\columnwidth]{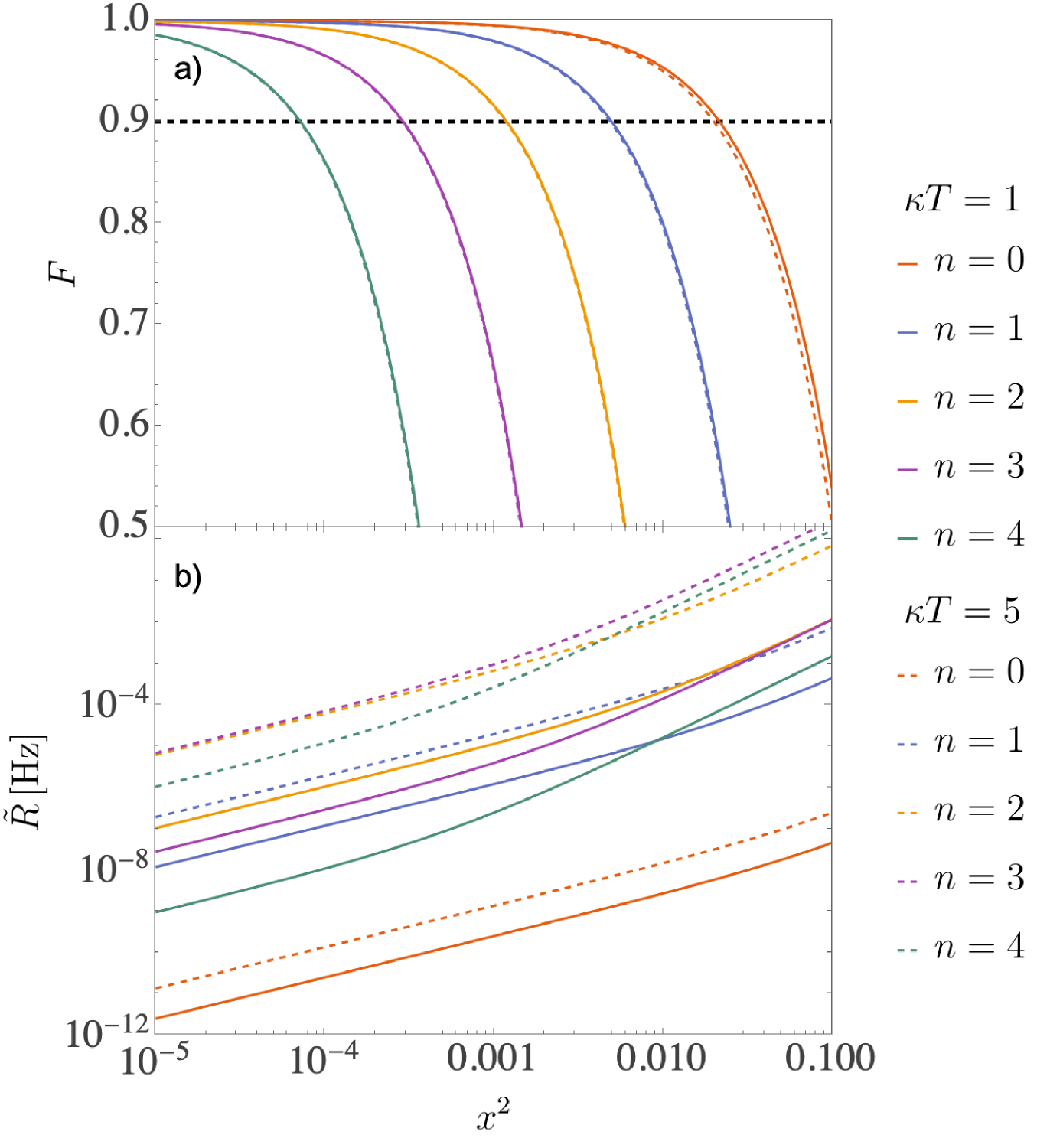}
    \caption{Fidelity and rate for continuously driven SPDC sources. Full (dashed) lines show the fidelity a) and un-multiplexed rate pr. quantum memory b) for different swap depths at different laser intensities (quantified with $x=1$ when the laser at threshold intensity) for $\kappa T=1$ ($\kappa T=5$). The dashed black line in a) shows 90\% fidelity, which is used as target fidelity. We have used $\eta_d=0.9$, $\eta_m=0.8$, $c=\unitfrac[2\cdot 10^5]{km}{s}$, $L_\text{att}=\unit[22]{km}$, $\kappa\mathcal{T}=100$ and $L=\unit[500]{km}$. For a) lines go in descending order in $n$. For b) the full (dashed) lines go in order 0, 4, 1, 3 and 2 (0, 1, 4, 2 and 3) at $x^2=10^{-5}$.}
    \label{fig:xplot}
\end{figure}

The fidelity and unmultiplexed rate are plotted as functions of the laser intensity in Fig. \ref{fig:xplot}. From the fidelity, we see that a higher swap depth leads to more stringent demands on the laser intensity due to higher impact of mode impurities. As expected, the rate increases with higher intensities. Since we want to obtain both a high rate and fidelity, we thus need to balance these effects.
\begin{figure}
    \includegraphics[width=\columnwidth]{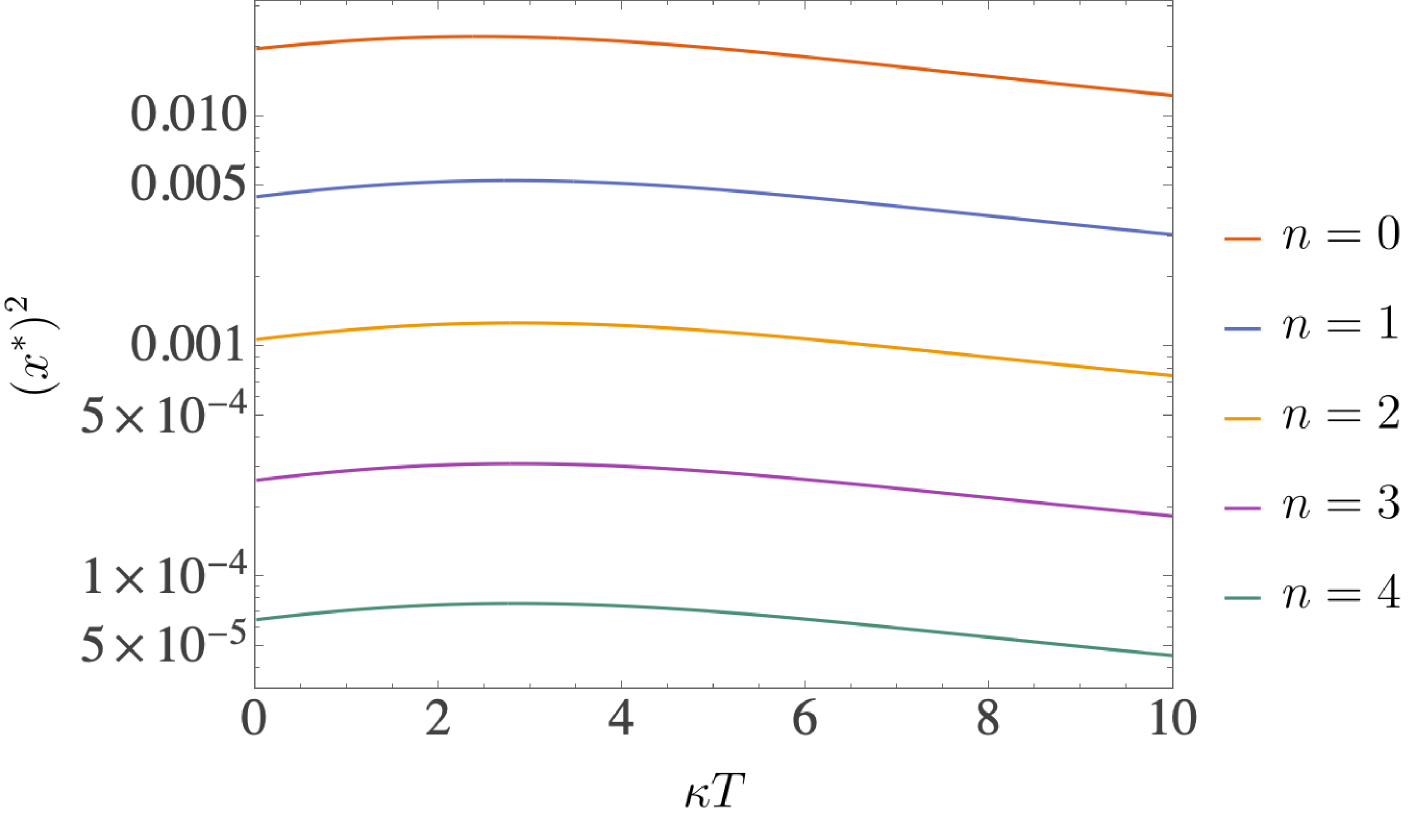}
    \caption{Optimal laser intensity for obtaining 90\% fidelity for continuous driving and different swap depths. We have used a target fidelity of $90\%$, $\eta_d=0.9$, $\eta_m=0.8$, $c=\unitfrac[2\cdot 10^5]{km}{s}$ and $L_\text{att}=\unit[22]{km}$. The lines go in descending order in $n$, as one moves up the $(x^*)^2$ axis.}
    \label{fig:target_x}
\end{figure}
\begin{figure}
    \includegraphics[width=\columnwidth]{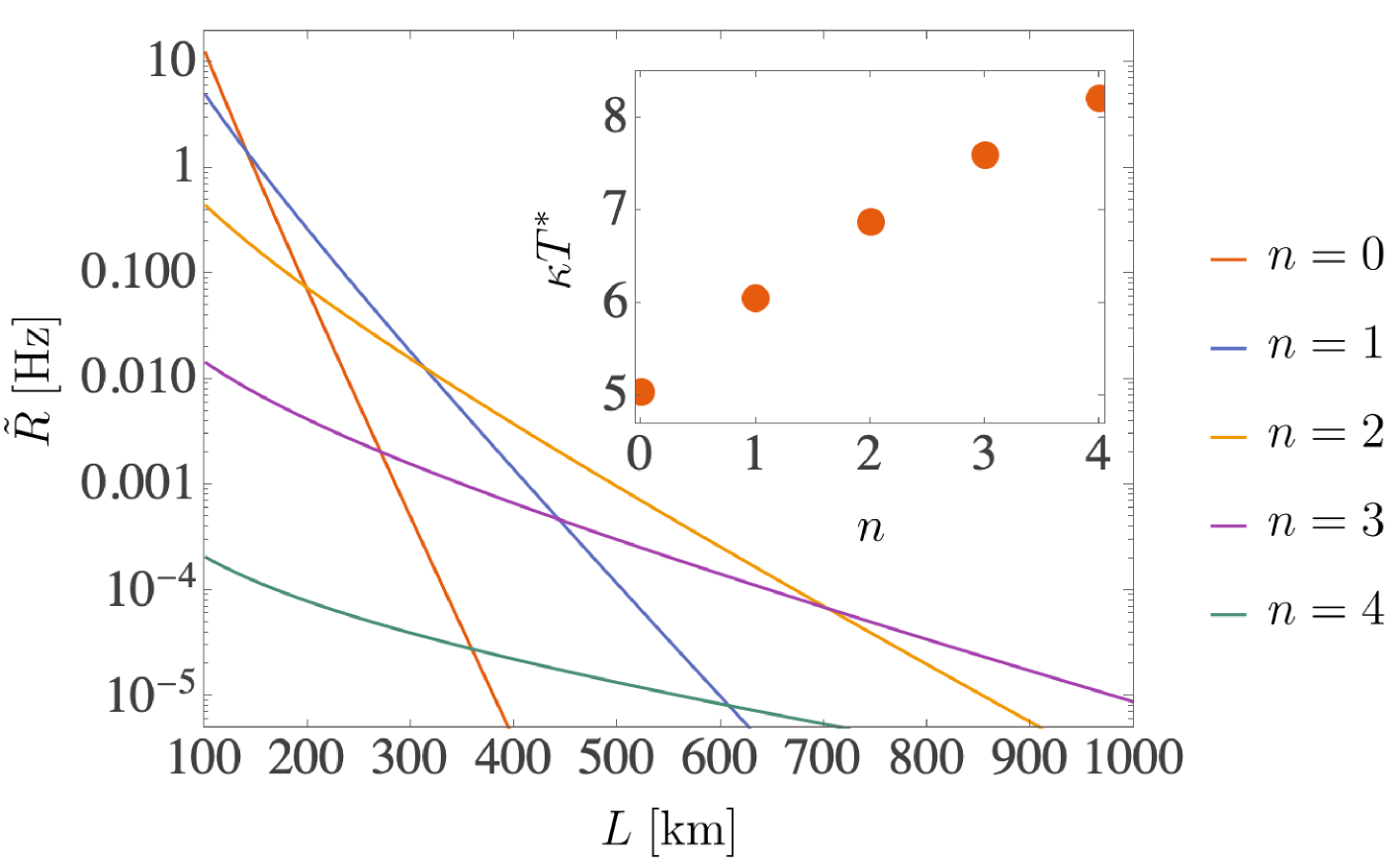}
    \caption{Rate per mode for CW driving for the different swap depths using the optimal interval $T^*$. In the inset, the optimal value  for the interval is shown. We have used a target fidelity of $90\%$, $\eta_d=0.9$, $\eta_m=0.8$, $c=\unitfrac[2\cdot 10^5]{km}{s}$, $L_\text{att}=\unit[22]{km}$ and $\kappa\mathcal{T}=100$. The lines go in descending order in $n$, as one moves up the $\tilde{R}$ axis at $L=\unit[100]{km}$}
    \label{fig:rates_n_cw}
\end{figure}

We find the highest allowed laser intensity to obtain a target fidelity of 90\%, and plot these in Fig. \ref{fig:target_x}. This laser power is then used to find the unmultiplexed rate pr. QM. We can factor the rate as $\tilde{R}(n,L,T) = f(n,L)g(n,T)$; hence, it is possible to find the optimal acceptance interval $T^*$ for the different swap depths regardless of the total distance. This is shown in the inset of Fig. \ref{fig:rates_n_cw}. Using this optimal acceptance interval for the different swap depths allows us to plot the unmultiplexed rate for the different swap depths over different combined distances, which is shown in Fig. \ref{fig:rates_n_cw}. As expected when increasing the distance, introducing entanglement swaps become advantageous. We summarize the results in Tab. \ref{tab:t*}.

\begin{table}
    \begin{tabular}{|c | c c c|}
        \hline $n$ & $\kappa T^*$ & $(x^*)^{2}$ & $L\, [\unit{km}]$\\\hline
        0 & 5.1 & $2.0\cdot 10^{-2}$ & 100 - 141\\
        1 & 6.1 & $4.5\cdot 10^{-3}$ & 141 - 313\\
        2 & 6.9 & $1.0\cdot 10^{-3}$ & 313 - 703\\
        3 & 7.6 & $2.3\cdot 10^{-4}$ & 703 - 1596\\
        4 & 8.2 & $5.4\cdot 10^{-4}$ & 1596 - \\
        \hline
    \end{tabular}
    \caption{The swap depths $n$ with the associated optimal acceptance interval $T^*$ and the range in which this swap depth delivers the highest rate (above $\unit[100]{km}$ and up to $n=4$). We have used a target fidelity of $90\%$, $\eta_d=0.9$, $\eta_m=0.8$ and $L_\text{att}=\unit[22]{km}$.}
    \label{tab:t*}
\end{table}

\section{Conclusion \& Discussion}\label{sec:conc}
We have investigated the influence of multimode nature of SPDC sources on a DLCZ type repeater. The impurity of the photon sources is shown to impact the possibly achievable rates in both cases considered in this study. The purity of the output state from the source leads to a deterioration of the fidelity of the final entangled pair the first entanglement swap. This leads to increasingly stringent limitations on the intensity of the driving pulse when adding more swaps.

When driving the source using a short pulse, the purity of the output state decreases when increasing the width of the driving pulse. Hence, when using longer pulses the constraints on the laser power become increasingly demanding when increasing the number of swaps. This ultimately leads to swap depths becoming unable to sustain a high fidelity of entangled pairs between the two ends. As a consequence, using long pulses in a DLCZ-type repeater has a limited advantage over direct transmission. The advantage is much clearer when employing narrow pulses, where high swap depths are possible. We also considered constraining the maximally allowed peak laser power. When increasing the swap depth, where the total laser power has to be decreased to avoid errors from two-photon emissions, we can decrease the pulse width, to obtain a more pure state. We find that even for realistic limitations on the laser power, we can operate closely to the ideal single-mode case.

When using continuous driving, we similarly see an impact of the full multimode nature of the SPDC source. To the lowest order in the driving strength the light from the source is in a pure single-photon state. However, as the laser power increases, the purity of the emitted state decreases due to multiple photons being emitted into other modes. The exact nature of the impurity and the probability of successfully swapping the entanglement depend on the size of the acceptance window. Using smaller acceptance windows limits the probability of getting successful swaps, however, any light stored in the QM will be in approximately pure state. Longer acceptance windows leads to higher chances of successfully swapping the entanglement, however, the state might be more impure both due to multiple photons and these photons occupying multiple modes both effects leads to a decrease in fidelity.

\begin{figure*}
    \begin{subfigure}{0.49\textwidth}
        \includegraphics[width=\textwidth]{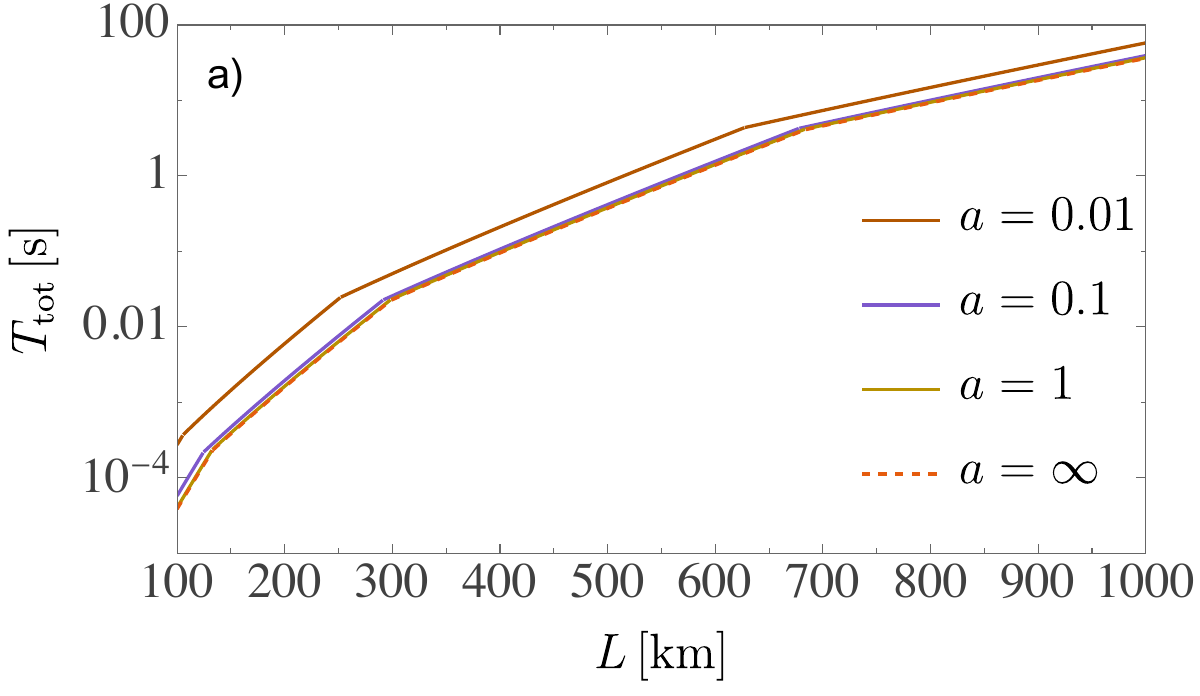}
    \end{subfigure}
    \begin{subfigure}{0.49\textwidth}
        \includegraphics[width=\textwidth]{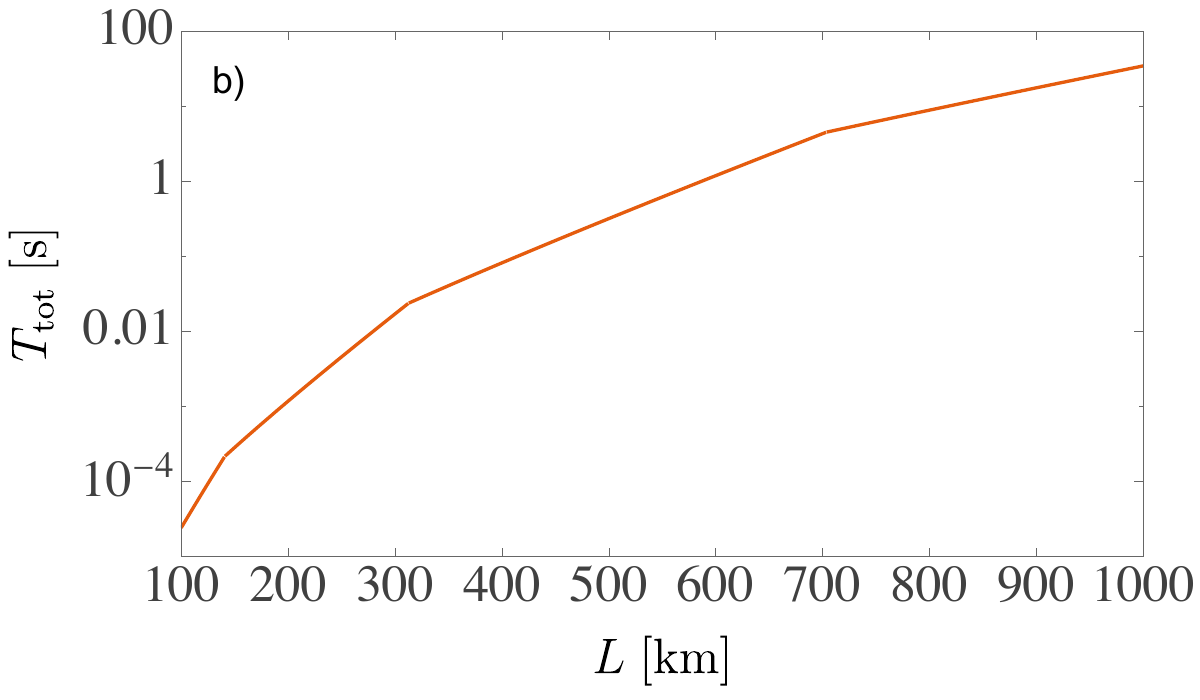}
    \end{subfigure}
    \caption{Time to generate entangled pair when including multiplexing for pulsed a) and CW b) driving of the SPDC. For both cases we have set 10 frequency and spatial multiplexing modes, and employ 32 memories. For the pulsed case we consider 10 temporal multiplexing modes, and for CW we have set $\kappa \mathcal{T} =100$. We stress that one should not compare the numbers directly between these two graphs, as it depends heavily on the numbers used in the two scenarios. We have used a target fidelity of 90\%, $\eta_d=0.9$, $\eta_m=0.8$, $c=\unitfrac[2\cdot 10^5]{km}{s}$ and $L_\text{att}=\unit[22]{km}$. For figure a) the lines correspond to descending value of $a$ as one go up the $T_\text{tot}$ axis.}
    \label{fig:time_total}
\end{figure*}

As highlighted in the beginning, the great advantage of atomic ensemble memories are their multiplexing abilities, allowing the rate to be scaled up dramatically. We can thus find the time to generate entanglement with multiplexing through $T_\text{tot}=(N_\text{mm} N_\text{mem} \tilde{R})^{-1}$, where $N_\text{mm}$ is the number of multiplexing modes pr. memory, and $N_\text{mem}$ is the number of memories used. These are plotted in Fig. \ref{fig:time_total}, where we have used 10 frequency, 10 spatial and 10 temporal multiplexing modes for the pulsed case, and 10 frequency and 10 spatial multiplexing modes with $\kappa \mathcal{T}=100$ for the continuous case, and we employ 32 memories in both cases. We see that entanglement generation times in the sub-second scale are possible at a distance of \unit[500]{km} to achieve with optimistic near-future efficiencies. 

We emphasize that these results are obtained for original DLCZ-repeater scheme. Multiple schemes have been developed since the original proposal, which achieve higher rates. In the future it would be interesting to perform similar investigations of these more advanced protocols. We believe that the results will be qualitatively similar, but the details will likely be different.

\section*{Acknowledgments}
The authors would like to thank Benedikt Tissot and Klaus Mølmer for useful discussions for the study.
This work was funded by the European Union's Horizon Europe research and innovation programme under grant agreement No. 101102140 – QIA Phase 1. We further acknowledge the support of Danmarks Grundforskningsfond (DNRF Grant No. 139,Hy-Q Center for Hybrid Quantum Networks).

\bibliographystyle{unsrt}
\bibliography{Merkur}
\newpage
\appendix
\onecolumngrid
\section{Brief summary of the model of the sources} \label{app:source}
We present a thorough investigation of SPDC sources in Ref \cite{hellebekCharacterizationMultimodeNature2024}. In this appendix we briefly summarize how the sources are modelled. 

We consider a non-linear crystal embedded in a cavity with a FWHM linewidth $\kappa$. A classical driving field is sent onto the crystal and we consider it to be unaffected by the cavity. This field will interact with the non-linear crystal according to the Hamiltonian 
\begin{align}
    H = \hbar\chi(t) (a_c b_c + a_c^\dagger b_c^\dagger),
\end{align} 
where $\chi(t)$ is the interaction strength between the field and the non-linear crystal with the time-dependence of the driving field and $a_c$ ($b_c$) is the annihilation operator of the signal (idler) cavity mode. The cavity modes described by $a_c$ and $b_c$ leak out into the output signal and idler fields $a(t)$ and $b(t)$. This leads to the input output relations for the annihilation operators (here shown for the signal field)
\begin{subequations}
    \begin{align}
        \dot{a}_c &= -\frac{\kappa}{2} a_c + \sqrt{\kappa} a_\text{in}(t) - i \chi(t) b_c^\dagger\\
        a(t) &= a_\text{in}(t) - \sqrt{\kappa} a_c,
    \end{align}
\end{subequations}
where $a_\text{in}(t)$ is the annihilation operator of the input operator. The differential equations can then be solved for a given driving field, and thus a given $\chi(t)$, since $\chi(t)$ is proportional to the incident field. The input-output relations reveal that if $\chi(t)> \kappa/2$, that the cavity field will increase exponentially. We thus define the threshold intensity as the incident intensity required to reach the threshold and normalize all incident intensities $I$ to the threshold intensity, i.e. $I/I_\text{thres} = 4\chi(t)^2/\kappa^2$.

\subsection{$P_1$ at threshold calculation}\label{app:P1}
We will in this appendix use the above parametrization to relate $P_1$ to the incident intensity relative to the threshold intensity. We recall the driving strength 
\begin{align}
    \chi(t) = \frac{x}{\sqrt{2\pi\sigma^2}}e^{-t^2/2\sigma^2},
\end{align}
where $x$ parametrizes the interaction strength.
To find $P_1$ to the lowest order in $x$, we have to compute $\int_\mathbb{R} dt\ev*{a^\dagger(t) a(t)}$ to the lowest order in $x$. With a time dependent interaction the probability $P_1$ is given by
\begin{align}
    P_1 = \int_{\mathbb{R}^2} dt d\tau \chi(t)\chi(\tau) e^{-\kappa\abs{t-\tau}}.
\end{align}
Inserting the Gaussian form of $\chi(t)$ and rewriting to make the inner integral up to $t$, the integral can be expressed in terms of an integral of an error function
\begin{align}
    P_1 = \frac{x^2}{\sqrt{2\pi\sigma^2}} \int_\mathbb{R} dt e^{-\pqty{\frac{t^2}{2\sigma^2}+\kappa t - \frac{\kappa^2\sigma^2}{2}}}\bqty{\erf\pqty{\frac{t}{\sqrt{2\sigma^2}}-\frac{\kappa\sigma}{\sqrt{2}}}+1}.
\end{align}
This integral can be found in Ref. \cite{ngTableIntegralsError1969}, leading to the result
\begin{align}
    P_1 = x^2e^{\kappa^2\sigma^2}\bqty{1-\erf\pqty{\kappa\sigma}}.
\end{align}
We now define $a=I_\text{peak}/I_\text{thres} = 2x^2/\pi\sigma^2\kappa^2$, we can solve for $x$ and obtain a value for $P_1$
\begin{align}
    P_{1,\text{max}} = a\frac{\pi \kappa^2\sigma^2}{2} e^{\kappa^2\sigma^2}\bqty{1-\erf\pqty{\kappa\sigma}}.
\end{align}
These are the lines plotted in Fig. \ref{fig:Threshold}.

\section{Density matrix for the pulsed case} \label{app:pulse}
We will in this appendix consider employing a pulsed driving field and describe the full density matrix, the recursive relations for swapping and their closed form solutions. We start with the general expression of the density matrix for the pulsed case
\begin{align}
    &\rho_n = c_{00,n} \ketbra{\emptyset}\ketbra{\emptyset} + \frac{c_{10,n}}{2}\sum_{\ell}w_\ell\pqty{\ketbra{1_\ell}\ketbra{\emptyset}+\ketbra{\emptyset}\ketbra{1_\ell}} + c_{11,n} \sum_{\ell,k} w_\ell w_k \ketbra{1_\ell}\ketbra{1_k}\nonumber \\\label{eq:dm_pulsed}
    & + \frac{c_{20,n}}{1+\varpi_1}\sum_{\ell,k\geqslant \ell} w_\ell w_k \pqty{\ketbra{1_\ell 1_k}\ketbra{\emptyset}+\ketbra{\emptyset}\ketbra{1_\ell 1_k}}
    + \frac{1}{2}\sum_{\ell}\bqty{A_{0,n}+A_{1,n} \Lambda_\ell}w_\ell^{2^n}\pqty{\ketbra{1_\ell}{\emptyset}\ketbra{\emptyset}{1_\ell}+\hc}\\\nonumber
    &+ \frac{B_{0,n}}{2} \sum_{\ell} w_\ell^{2^n}\sqrt{\frac{2}{1+\varpi_1}}\bqty{\sqrt{2}w_\ell\pqty{\ketbra{1_\ell}{\emptyset}\ketbra{1_\ell}{2_\ell}+\ketbra{2_\ell}{1_\ell}\ketbra{\emptyset}{1_\ell}}+\sum_{k\neq \ell} w_k \pqty{\ketbra{1_\ell}{\emptyset}\ketbra{1_k}{1_k1_\ell}+\ketbra{1_k1_\ell}{1_k}\ketbra{\emptyset}{1_\ell}}+\hc},
\end{align}
%NOTE REDEFINITION OF LAMBDA K

where $\Lambda_\ell = \sqrt{\frac{2}{1+\varpi_1}}(1+w_\ell)$. The initial conditions for the coefficients are 

\begin{subequations}
\begin{align}
    c_{00,0}&=0,& c_{10,0}&=\frac{1-P_1}{1+P_1\pqty{1+\varpi_1}},&
    c_{11,0}&=\frac{P_1}{1+P_1\pqty{1+\varpi_1}},& c_{20,0}&=\frac{P_1\pqty{1+\varpi_1}}{1+P_1\pqty{1+\varpi_1}}, \\
     A_{0,0} &= \frac{1-P_1}{1+P_1\pqty{1+\varpi_1}},&
    A_{1,0}&=0,&
    B_{0,0}&=\frac{P_1\sqrt{\pqty{1+\varpi_1}/2}}{1+P_1\pqty{1+\varpi_1}}. &&
\end{align}
\end{subequations}
We can find the recursive equations for the density matrix after entanglement swapping from Eq. \eqref{eq:eswap}. We will however ignore the renormalization, i.e. the denominator, as the equations become much easier to combine. Hence, $\rho_{n}$ is not normalised in the appendix, instead the trace yields the product $\Tr[\rho_n]=\prod_{i=1}^n(\mathbf{P}_{n-i}/2)^{2^i}$.

The recursive equations will form two sets, of interconnected equations. One involving the diagonal elements, i.e. $c_{00}$, $c_{10}$, $c_{11}$ and $c_{20}$, and one involving the off-diagonal elements, i.e. $A_{0}$, $A_{1}$ and $B_0$.

We start by considering the diagonal elements, we obtain the recursive equations
\begin{subequations}\label{eq:rec}
\begin{align}
    c_{00,n+1}&= \eta_2\bqty{\frac{c_{10,n}}{2}+c_{20,n}\pqty{1-\eta_2}}\bqty{c_{00,n}+\frac{c_{10,n}\pqty{1-\eta_2}}{2}+\frac{c_{20,n}\pqty{1-\eta_2}^2}{2}},\\
    c_{10,n+1} &= \eta_2\qty{\bqty{\frac{c_{10,n}}{2}+c_{20,n}\pqty{1-\eta_2}}\bqty{\frac{c_{10,n}}{2}+c_{11,n}\pqty{1-\eta_2}}+c_{11,n}\bqty{c_{00,n}+\frac{c_{10,n}\pqty{1-\eta_2}}{2}+\frac{c_{20,n}\pqty{1-\eta_2}^2}{2}}},\hspace{-1ex}\\
    c_{11,n+1} &= \eta_2 \frac{c_{10,n}}{2}c_{11,n},\\
    c_{20,n+1} &= \eta_2 \frac{c_{10,n}}{2}\frac{c_{20,n}}{2},
\end{align}
\end{subequations}
where $\eta_2=\eta_m\eta_d$.
To find closed forms for the recursive equations, we expand the coefficients and their recursive equations into orders of $P_1$, and solve each separately. Thus, we find closed forms of the equations up to first order in $P_1$. As emissions of more than two photon pairs from the source are discarded, terms of the order $P_1^2$ will not be fully accounted for. Thus, our results will only be valid up to first order in $P_1$, regardless of expansion here
\begin{subequations}
\begin{align}\label{eq:RecEqc00}
    c_{00,n} &= \pqty{\frac{\eta_2}{4}}^{2^n-1}(1-\eta_2)(2^n-1)\qty{
        1-P_1\bqty{\pqty{1+\varpi_1}\pqty{1-\eta_2}+2^{n}\frac{4-3\varpi_1+2\eta_2+6\varpi_1\eta_2}{3}-2^{2n+1}\frac{1-\eta_2}{3}}
    }\\
    c_{10,n} &= \pqty{\frac{\eta_2}{4}}^{2^n-1}\qty{
        1-P_1\bqty{2\pqty{1+\varpi_1}\pqty{1-\eta_2}+2^{n}\pqty{2-\varpi_1+2\varpi_1\eta_2}-2^{2n+1}\pqty{1-\eta_2}}
    }\\
    c_{11,n} &= 2^n\pqty{\frac{\eta_2}{4}}^{2^n-1} P_1\\
    c_{20,n} &= \pqty{\frac{\eta_2}{4}}^{2^n-1}\pqty{1+\varpi_1} P_1.
\end{align}
\end{subequations}
% \end{widetext}
We now do the same for the off-diagonal elements. The recursive equations are
\begin{subequations}
\begin{align}
    A_{0,n+1} &= \frac{\eta_2}{4}A_{0,n}^2,\\
    A_{1,n+1} &= \frac{\eta_2}{2}A_{0,n}\bqty{A_{1.n}+\pqty{1-\eta_2}B_{n,0}},\\
    B_{0,n+1} &= \frac{\eta_2}{4}A_{0,n}B_{n,0}.
\end{align}
\end{subequations}

These recursive equations can be solved directly. We will only give the answer expanded to first order of $P_1$, as higher order contributions (e.g. three photon emissions) are not included in the original state
\begin{subequations}
\begin{align}
    A_{0,n} &= \pqty{\frac{\eta_2}{4}}^{2^n-1}\bqty{1-2^n\pqty{2+\varpi_1}P_1},\\
    A_{1,n} &= \pqty{\frac{\eta_2}{4}}^{2^n-1}(2^n-1)(1-\eta_2)\sqrt{\frac{1+\varpi_1}{2}}P_1,\\
    B_{0,n} &= \pqty{\frac{\eta_2}{4}}^{2^n-1}\sqrt{\frac{1+\varpi_1}{2}}P_1.
\end{align}
\end{subequations}
All calculations here can be found in the code base provided in Ref. \cite{hellebekCodeBaseAvailable}.

\section{Density matrix for continuous driving} \label{app:CW}
In this appendix, we describe the density matrix when employing a CW driving laser, and the recursive equations for the coefficients in the density matrix. As noted in the main text, the density matrix depends on the click time, $t_c$, and the acceptance window, $T$. We assume $t_c=0$, and we will not write the dependence on $T$ for ease of reading. A different detection time will only shift the density matrix to be symmetric around $t_c$ (ignoring for simplicity  edge effects where the click time is close to the end of the time interval of the driving).

We will consider the density matrix
% \begin{widetext}
\begin{align}
    \rho_{n} = &c_{00,n}\ketbra{\emptyset}\ketbra{\emptyset}
    + \frac{c_{10,n}}{2}\int_{-\frac{T}{2}}^{T/2} d\qty{t} \rho_{10}(\qty{t})\pqty{\ketbra{t_1}{t_2}\ketbra{\emptyset}+\ketbra{\emptyset}\ketbra{t_1}{t_2}}\nonumber\\
    &+ \frac{1}{2}\int_{-\frac{T}{2}}^{T/2} d\qty{t} \bqty{A_{0,n}\rho_{A0}(\qty{t})+A_{1,n}\rho_{A1}(\qty{t})+A_{2,n}\rho_{A2}(\qty{t})}\pqty{\ketbra{t_1}{\emptyset}\ketbra{\emptyset}{t_2}+\text{h.c}}\\\nonumber
    &+ c_{11,n}\int_{-\frac{T}{2}}^{T/2} d\qty{t} \rho_{11}(\qty{t})\ketbra{t_1}{t_3}\ketbra{t_2}{t_4}
    + \frac{c_{20,n}}{8}\int_{-\frac{T}{2}}^{T/2} d\qty{t} \rho_{20|T}(\qty{t})\pqty{\ketbra{t_1t_2}{t_3t_4}\ketbra{\emptyset}+\ketbra{\emptyset}\ketbra{t_1t_2}{t_3t_4}}\\\nonumber
    &+\frac{B_{0,n}}{4}\int_{-\frac{T}{2}}^{T/2} d\qty{t} \rho_{B0}(\qty{t})\pqty{\ketbra{t_1t_2}{t_3}\ketbra{\emptyset}{t_4}+\ketbra{\emptyset}{t_4}\ketbra{t_1t_2}{t_3}+\text{h.c.}},
\end{align}
where $\qty{t}$ includes all time-coordinates. We take the functions $\rho_{ij}$ to be normalized appropriately.
To find the functions and initial values for the coefficients, i.e. the density matrix after entanglement generation, we will follow the methods followed developed in Ref. \cite{hellebekCharacterizationMultimodeNature2024}, where the two- and four-time correlation functions after the detection are used to describe the density matrix, up to second order in $x=2\chi/\kappa$. We write down the useful combinations of the pre-detection two-time correlation functions up to second order in $x$, where the idler photon is always taken at $t_c=0$, i.e. $a_i=a_i(0)$ etc.
\begin{align}
    \frac{\ev*{a_s^\dagger(t)a_i^\dagger}\ev*{a_s(t')a_i}}{\ev*{a_i^\dagger a_i}} &= \frac{\kappa}{2}e^{-\frac{\kappa}{2}\pqty{\abs{t}+\abs{t'}}}\qty{1+\frac{x^2}{2}\bqty{\pqty{1+\frac{\kappa\abs{t}}{2}}^2+\pqty{1+\frac{\kappa\abs{t'}}{2}}^2}}\\
    \ev*{a_s^\dagger(t)a_s(t')} &= \frac{\kappa}{2} x^2 e^{-\frac{\kappa}{2}\abs{t-t'}}\pqty{1+\frac{\kappa\abs{t-t'}}{2}}.
\end{align}
We will, as an example, describe how to calculate two of the elements in the density matrix. Consider the element corresponding to two photons in one quantum memory. Only one four-time correlation function contributes up to second order in $x$ 
\begin{align}
    &\frac{c_{20,0}}{8}\bqty{\rho_{n,20}(t_1,t_2,t_3,t_4)+ (t_1\leftrightarrow t_2) + (t_3\leftrightarrow t_4) + (t_1\leftrightarrow t_2,\, t_3\leftrightarrow t_4)}
    =\ev*{a_s^\dagger(t_4)a_s^\dagger(t_3)a_s(t_2)a_s(t_1)}\nonumber\\\nonumber 
    &= \frac{\ev*{a_s^\dagger(t_4)a_s^\dagger(t_3)a_s(t_2)a_s(t_1)b_+^\dagger b_+}}{\ev*{b_+^\dagger b_+}} 
    = \frac{\ev*{a_s^\dagger(t_4)a_s(t_2)}\ev*{a_s^\dagger(t_3)a_i^\dagger}\ev*{a_s(t_1)a_i}}{2\ev*{a_i^\dagger a_i}} + (t_1\leftrightarrow t_2) + (t_3\leftrightarrow t_4) + (t_1\leftrightarrow t_2,\, t_3\leftrightarrow t_4) \\ 
    &\to\, \frac{c_{20,0}}{4}\rho_{n,20}(t_1,t_2,t_3,t_4) = \ev*{a_s^\dagger(t_4)a_s(t_2)}\frac{\ev*{a_s^\dagger(t_3)a_i^\dagger}\ev*{a_s(t_1)a_i}}{\ev*{a_i^\dagger a_i}},
\end{align}
where $(t_i\leftrightarrow t_j)$ indicates the first term where $t_i$ and $t_j$ are swapped time coordinates. 

To find the single-photon density matrix elements, we have to consider the contribution of two-photon density matrix elements to the two-time correlation functions. In effect, this means that we should subtract appropriate four-time correlation functions from the single photon correlation function. We will as an example consider the density matrix element with one photon in the one quantum memory
\begin{align}
    \frac{c_{10,0}}{2}\rho_{10}(\qty{t}) =&\frac{\ev*{a_s^\dagger(t_2)a_s(t_1)b_+^\dagger b_+}}{\ev*{b_+^\dagger b_+}}\\\nonumber &- \int_{-\frac{T}{2}}^\frac{T}{2} d\tau \qty{c_{11,0}\rho_{11}(t_1,\tau,t_2,\tau)+\frac{c_{20,0}}{8}\bqty{\rho_{20}(t_1,\tau,t_2,\tau)+\rho_{20}(t_1,\tau,\tau,t_2)+\rho_{20}(\tau,t_1,t_2,\tau)+\rho_{20}(\tau,t_1,\tau,t_2)}}\\\nonumber
    =&\frac{\ev*{a_s^\dagger(t_2)a_i^\dagger}\ev*{a_s(t_1)a_i}}{\ev*{a_i^\dagger a_i}}\pqty{\frac{1}{2}-\int_{-\frac{T}{2}}^\frac{T}{2}d\tau\ev*{a_s^\dagger(\tau)a_s(\tau)}} +\ev*{a_s^\dagger(t_2)a_s(t_1)}\pqty{1-\int_{-\frac{T}{2}}^\frac{T}{2} d\tau\frac{\abs{\ev*{a_s(\tau)a_i}}^2}{\ev*{a_i^\dagger a_i}}}\\\nonumber
    &-\int_{-\frac{T}{2}}^\frac{T}{2} d\tau \frac{\ev*{a_s^\dagger(\tau)a_s(t_1)}\ev*{a_s^\dagger(t_2)a_i^\dagger}\ev*{a_s(\tau)a_i}+\ev*{a_s^\dagger(t_2)a_s(\tau)}\ev*{a_s^\dagger(\tau)a_i^\dagger}\ev*{a_s(t_1)a_i}}{2\ev*{a_i^\dagger a_i}}. 
\end{align}
To write it down properly, we split the coefficient and function into terms attached to a specific order in $x$, i.e. $c_{10,0}\rho_{10}(\qty{t}) = c_{10,0,0}\rho_{10,0}(\qty{t})+c_{10,0,2}\rho_{10,2}(\qty{t})$. Following these steps for all the coefficients we can find the values of the coefficients after the entanglement generation. We will normalize it such that the density matrix functions integrate to 1. Starting with the diagonal elements, i.e. $c_{00},\, c_{10}\, c_{11}$ and $c_{20}$
\begin{subequations}
    \begin{align}
        c_{00,0,0} &= e^{-\frac{\kappa T}{2}} && c_{00,0,2} =  -\frac{x^2}{4}\qty{10\pqty{1-e^{-\frac{\kappa T}{2}}}+\kappa T\bqty{2-e^{\frac{\kappa T}{2}}\pqty{3+\frac{\kappa T}{4}}}}  \\
        c_{10,0,0}&= 1-e^{-\frac{\kappa T}{2}}, &&
        c_{10,0,2}= -\frac{x^2}{4}\bqty{2\kappa T-e^{-\frac{\kappa T}{2}}\pqty{4+5\kappa T}+e^{-\kappa T}\pqty{4+\kappa T}},\\
        c_{11,0} &=  \frac{x^2}{2} \kappa T\pqty{1-e^{-\frac{\kappa T}{2}}}, &&
        c_{20,0} = \frac{x^2}{4}\bqty{2\pqty{5+\kappa T}-e^{-\frac{\kappa T}{2}}\pqty{14+6\kappa T+\frac{\kappa^2 T^2}{4}}+e^{-\kappa T}\pqty{4+\kappa T}}.
    \end{align}
\end{subequations}
As the swapping method is identical for the two cases and since the diagonal matrix coefficients combine in the same way, the recursive equations are the same for the two cases. Hence, we will use the recursive equations for the pulsed case (see. Eq. \eqref{eq:rec}). Closed forms for the coefficients can be found using an analytical software like Mathematica, and this is provided in the code base \cite{hellebekCodeBaseAvailable}. We can also find the density matrix functions for these elements
\begin{subequations}
    \begin{align}
    \rho_{10,0}(\qty{t}) &= \frac{\kappa e^{-\frac{\kappa}{2}\pqty{\abs{t_1}+\abs{t_2}}}}{2\pqty{1-e^{-\frac{\kappa T}{2}}}}\\
    \rho_{10,2}(\qty{t}) &= \frac{2\kappa e^{-\frac{\kappa T}{2}}e^{-\frac{\kappa\abs{t_1-t_2}}{2}}\pqty{1+\frac{\kappa \abs{t_1-t_2}}{2}}}{2\kappa T-e^{-\frac{\kappa T}{2}}\pqty{4+5\kappa T}+e^{-\kappa T}\pqty{4+\kappa T}}\nonumber\\
            &\quad+\frac{\kappa e^{-\frac{\kappa}{2}\pqty{\abs{t_1}+\abs{t_2}}}/2}{2\kappa T-e^{-\frac{\kappa T}{2}}\pqty{4+5\kappa T}+e^{-\kappa T}\pqty{4+\kappa T}}
            \bigg\{
                4+2\kappa T +\frac{\kappa \pqty{\abs{t_1}+\abs{t_2}}}{2}\pqty{1-e^{-\frac{\kappa T}{2}}}\\\nonumber
                &\quad+ \frac{e^{-\frac{\kappa T}{2}}}{2}\bqty{
                    \frac{\kappa\pqty{\abs{t_1}e^{\kappa \abs{t_1}}+\abs{t_2}e^{\kappa \abs{t_2}}}}{2}
                    -\pqty{1+ \frac{e^{\kappa \abs{t_1}}+ e^{\kappa \abs{t_2}}}{2}}\pqty{6+\kappa T}}
            \bigg\}\\
    \rho_{11}(\qty{t}) &= \frac{\kappa \bqty{e^{-\frac{\kappa}{2}\pqty{\abs{t_1}+\abs{t_3}+\abs{t_2-t_4}}}\pqty{1+\frac{\kappa\abs{t_2-t_4}}{2}}+e^{-\frac{\kappa}{2}\pqty{\abs{t_2}+\abs{t_4}+\abs{t_1-t_3}}}\pqty{1+\frac{\kappa\abs{t_1-t_3}}{2}}}}{4 T\pqty{1-e^{-\frac{\kappa T}{2}}}}\\
    \rho_{20}(\qty{t}) &= \frac{4\kappa^2e^{-\frac{\kappa}{2}\pqty{\abs{t_1}+\abs{t_3}+\abs{t_2-t_4}}}\pqty{1+\frac{\kappa\abs{t_2-t_4}}{2}}}{2\pqty{5+\kappa T}-e^{-\frac{\kappa T}{2}}\pqty{14+6\kappa T+\frac{\kappa^2 T^2}{4}}+e^{-\kappa T}\pqty{4+\kappa T}}.
\end{align}
\end{subequations}
We now turn to the off-diagonal density matrix elements. The functions are normalised, such that they don't interfere in the swapping process. The coefficients then become
\begin{subequations}
    \begin{align}
        A_{0,0} &= \pqty{1-e^{-\frac{\kappa T}{2}}}\pqty{1-\frac{x^2}{4} \kappa T}\\
        A_{1,0} &= \frac{x^2}{4} \bqty{10\pqty{1-e^{-\frac{\kappa T}{2}}} - \kappa T e^{-\frac{\kappa T}{2}}\pqty{3-\kappa T}}\\
        A_{2,0} &= -\frac{x^2}{4} \bqty{5 \pqty{1-e^{-\frac{\kappa T}{2}}}  - e^{-\frac{\kappa T}{2}}\pqty{2+2\kappa T +\frac{\kappa ^2 T^2}{8}}+ e^{-\kappa T}\frac{4+\kappa T}{2}}\\
        B_{0,0} &= \frac{x^2}{4}\bqty{\pqty{5+\kappa T} \pqty{1-e^{-\frac{\kappa T}{2}}}-e^{-\kappa T}\pqty{2+2\kappa T+\frac{\kappa^2 T^2}{8}}+e^{-\kappa T}\frac{4+\kappa T}{2}}.
    \end{align}
\end{subequations}
As the density matrix for the off-diagonal elements is slightly different in the CW case compared to the pulsed case, the recursive equations for the coefficient are also slightly different
\begin{subequations}
    \begin{align}
        A_{0,n+1} &= \eta_2 \frac{A_{0,n}}{4}\bqty{A_{0,n}+A_{1,n}+A_{2,n}+2(1-\eta_2)B_{0,n}}\\
        A_{1,n+1} &= \eta_2 \frac{A_{0,n}A_{1,n}}{4}\\
        A_{2,n+1} &= \eta_2 \frac{A_{0,n}A_{2,n}}{4}\\
        B_{0,n+1} &= \eta_2 \frac{A_{0,n}B_{0,n}}{4}.
    \end{align}
\end{subequations}
The recursive equations can again be solved using analytical software, and is again provided in the code base (REF).
We also give the off-diagonal matrix functions
\begin{subequations}
\begin{align}
    \rho_{A0}(\qty{t}) &= \frac{\frac{\kappa}{2}e^{-\frac{\kappa}{2}\pqty{\abs{t_1}+\abs{t_2}}}}{1-e^{-\frac{\kappa T}{2}}}\\
    \rho_{A1}(\qty{t}) &= \frac{\kappa e^{-\frac{\kappa}{2}\pqty{\abs{t_1}+\abs{t_2}}}\bqty{\pqty{1+\frac{\kappa \abs{t_1}}{2}}^2+\pqty{1+\frac{\kappa \abs{t_2}}{2}}^2}}{10\pqty{1-e^{-\frac{\kappa T}{2}}} - \kappa T e^{-\frac{\kappa T}{2}}\pqty{3-\kappa T}}\\
    \rho_{A2}(\qty{t}) &= \frac{\frac{\kappa}{2} e^{-\frac{\kappa}{2}\pqty{\abs{t_1}+\abs{t_2}}}}{5 - e^{-\frac{\kappa T}{2}}\pqty{7+2\kappa T +\frac{\kappa ^2 T^2}{8}}+ e^{-\kappa T}\frac{4+\kappa T}{2}}\bigg\{6+\pqty{3-e^{-\frac{\kappa T}{2}}}\frac{\kappa\pqty{\abs{t_1}+\abs{t_2}}}{2}+\frac{\kappa^2\pqty{\abs{t_1}^2+\abs{t_2}^2}}{4}\\\nonumber
    &\qquad\qquad\qquad\qquad\qquad-e^{-\frac{\kappa T}{2}}\pqty{3+\frac{\kappa T}{2}}\pqty{1+\frac{e^{\kappa\abs{t_1}}+e^{\kappa\abs{t_2}}}{2}}+e^{-\frac{\kappa T}{2}}\frac{\kappa \pqty{\abs{t_1}e^{\kappa\abs{t_1}}+\abs{t_2}e^{\kappa\abs{t_2}}}}{2}\bigg\}
    \\
    \rho_{B0,0}(\qty{t}) &= \frac{\kappa^2e^{-\frac{\kappa}{2}\pqty{\abs{t_1}+\abs{t_4}+\abs{t_2-t_3}}}\pqty{1+\frac{\kappa\abs{t_2-t_3}}{2}}}{\pqty{5+\kappa T}-e^{-\frac{\kappa T}{2}}\pqty{7+3\kappa T+\frac{\kappa^2 T^2}{8}}+e^{-\kappa T}\frac{4+\kappa T}{2}}.
\end{align}
\end{subequations}

\end{document}